%% file: main.tex
\documentclass[
  journal=pasa,
  manuscript=research-paper, 
  year=2025,
  volume=YY,
]{cup-journal}

\usepackage{amsmath}
\usepackage[nopatch]{microtype}
\usepackage{booktabs}

\usepackage{lscape}

\usepackage{float,xcolor}
\usepackage[T1]{fontenc}

\usepackage{graphicx}	
\usepackage{amsmath}	
\usepackage{float}
 \usepackage{amssymb}	
\usepackage{geometry}
\usepackage{hyphenat}
\usepackage{mwe}
\usepackage{blindtext}
\usepackage{wrapfig}
\usepackage[figuresright]{rotating}
\usepackage{gensymb}
\usepackage{natbib}
\usepackage{tablefootnote}

\usepackage{caption}
\DeclareCaptionFormat{bold}{\textbf{#1 #2 }#3}
\DeclareCaptionLabelSeparator{dot}{.}
\usepackage{svg}
\usepackage{array}
\usepackage{booktabs}

\newcommand{\Msun}{\mbox{\,$\rm M_{\odot}$}}
\newcommand{\Lsun}{\mbox{\,$\rm L_{\odot}$}}
\newcommand{\Lsed}{\mbox{\,$\rm L_{SED}$}}

\newcommand{\Teff}{$T_{\rm eff}$}

\newcommand{\logg}{$\log g$}
\newcommand{\mv}{$\xi_{\rm t}$}

\newcommand{\cfe}{$\textrm{[C/Fe]}$}
\newcommand{\ofe}{$\textrm{[O/Fe]}$}

\newcommand{\sfe}{$\textrm{[s/Fe]}$}

\newcommand{\lsfe}{$\textrm{[ls/Fe]}$}
\newcommand{\hsfe}{$\textrm{[hs/Fe]}$}
\newcommand{\hsls}{$\textrm{[hs/ls]}$}
\newcommand{\feh}{$\textrm{[Fe/H]}$}
\newcommand{\xfe}{$\textrm{[X/Fe]}$}
\newcommand{\xfelte}{$\xfe_{\rm LTE}$}
\newcommand{\xfenlte}{$\xfe_{\rm NLTE}$}

\newcommand{\kms}{km\,s$^{-1}$}

\newcommand{\sprocess}{\textit{s}-process }
\newcommand{\rprocess}{\textit{r}-process }
\newcommand{\iprocess}{\textit{i}-process }
\newcommand{\logepsilon}{$\log\varepsilon$}
\newcommand{\logepsilonsun}{$\log\varepsilon_{\odot}$}
\newcommand{\sigmatotal}{$\sigma_{\rm tot}$}
\newcommand{\sigmalinetoline}{$\sigma_{\rm |121|}$}

\newcommand{\pbfe}{$\textrm{[Pb/Fe]}$}
\newcommand{\pbls}{$\textrm{[Pb/ls]}$}
\newcommand{\pbhs}{$\textrm{[Pb/hs]}$}
\newcommand{\pbIfe}{$\textrm{[Pb I/Fe]}$}
\newcommand{\pbIIfe}{$\textrm{[Pb II/Fe]}$}

\usepackage{subcaption}
\usepackage{setspace,hyperref}
\hypersetup{colorlinks=true,
            linkcolor=blue,
            urlcolor=blue,
            linktoc=all,
            citecolor=blue}

\title{\sprocess Enriched Post-AGB Star J003643.94$-$723722.1 in the SMC with an Extreme C/O Ratio and the First Precise Detection of Lead}

\author{Meghna Menon}
\affiliation{School of Mathematical and Physical Sciences, Macquarie University, Balaclava Road, Sydney, NSW 2109, Australia}
\alsoaffiliation{Astrophysics and Space Technologies Research Centre, Macquarie University, Balaclava Road, Sydney, NSW 2109, Australia}
\email[Meghna Menon]{meghnamukesh.menon1@students.mq.edu.au}

\author{Devika Kamath}
\affiliation{School of Mathematical and Physical Sciences, Macquarie University, Balaclava Road, Sydney, NSW 2109, Australia}
\alsoaffiliation{Astrophysics and Space Technologies Research Centre, Macquarie University, Balaclava Road, Sydney, NSW 2109, Australia}
\alsoaffiliation{INAF, Observatory of Rome, Via Frascati 33, I-00077 Monte Porzio Catone (RM), Italy}

\author{Maksym Mohorian}
\affiliation{School of Mathematical and Physical Sciences, Macquarie University, Balaclava Road, Sydney, NSW 2109, Australia}
\alsoaffiliation{Astrophysics and Space Technologies Research Centre, Macquarie University, Balaclava Road, Sydney, NSW 2109, Australia}

\author{Anish M. Amarsi}
\affiliation{Theoretical Astrophysics, Department of Physics and Astronomy, Uppsala University, Box 516, SE-751 20 Uppsala, Sweden}

\author{Diego Vescovi}
\affiliation{INAF – Osservatorio astronomico d’Abruzzo, Via Mentore Maggini, 64100 Teramo, Italy}
\alsoaffiliation{INFN, Sezione di Perugia, Via A. Pascoli, 06123 Perugia, Italy}

\author{Sergio Cristallo}
\affiliation{INAF – Osservatorio astronomico d’Abruzzo, Via Mentore Maggini, 64100 Teramo, Italy}
\alsoaffiliation{INFN, Sezione di Perugia, Via A. Pascoli, 06123 Perugia, Italy}

\author{Amanda Karakas}
\affiliation{School of Physics and Astronomy, Monash University, VIC 3800, Australia}
\alsoaffiliation{ARC Centre of Excellence for All Sky Astrophysics in 3 Dimensions (ASTRO 3D), Clayton 3800, Australia}

\author{Hans Van Winckel}
\affiliation{Instituut voor Sterrenkunde, K.U.Leuven, Celestijnenlaan 200D bus 2401, B-3001, Leuven, Belgium}

\author{Paolo Ventura}
\affiliation{INAF, Observatory of Rome, Via Frascati 33, I-00077 Monte Porzio Catone (RM), Italy}

\received {dd Mmm YYYY}
\revised  {dd Mmm YYYY}
\accepted {dd Mmm YYYY}
\published{dd Mmm YYYY}


\begin{document}

\begin{abstract}
Post-asymptotic giant branch (post-AGB) stars are exquisite tracers of \sprocess nucleosynthesis, preserving the surface chemical signatures of their AGB evolution. The increasing chemical diversity observed among them challenges current nucleosynthesis models and motivates detailed case studies. In this study, we present a comprehensive abundance analysis of J003643.94$-$723722.1 (J003643), a single post-AGB star in the Small Magellanic Cloud (SMC). High-resolution UVES/VLT spectra analysed with E-iSpec reveal a C/O ratio of 16.21 and an \sfe\,= 2.09$\pm$0.20 dex. In this study, we also report the first direct detection of lead in a post-AGB star via the Pb II 5608.853~\AA\ line, with a derived \pbfe\,= 3.18 dex. Comparison with a comprehensive and appropriate sample of post-AGB stars across the Galaxy, Large Magellanic Cloud (LMC) and SMC shows that J003643 has a relatively high C/O ratio, far exceeding the typical range of $\sim$1–3. J003643's \cfe\ (1.33$\pm$0.14 dex) and \sfe\ (2.09$\pm$0.20 dex) are consistent with expectations from standard third dredge-up (TDU) enrichment. However, its \ofe\ (–0.08$\pm$0.20 dex) is significantly lower than that of the comparative sample with similar \cfe\ and \feh, which typically show \ofe\ between 0.5 and 1.0 dex. This relatively low \ofe, along with an [$\alpha$/Fe]$\,\approx0$ dex of J003643, is consistent with the chemical evolution of the SMC at $\feh\,\approx-1$ dex, in contrast to the oxygen-enhanced Galactic and LMC trend at $\feh\,\approx-1$ dex. This indicates that J003643’s high C/O ratio primarily results from its intrinsic oxygen deficiency rather than from an unusually high carbon enhancement. To better understand the CNO, alpha, Fe-peak, and heavy element nucleosynthesis, we compared J003643's abundance pattern with yields from three stellar evolutionary codes: ATON, MONASH, and FRUITY, the latter two incorporating post-processing nucleosynthesis. While these models reproduce the majority of elemental abundances, they significantly underpredict the Pb abundance, highlighting a persistent gap in our understanding of heavy element production in AGB stars. J003643 represents the second \sprocess enriched single post-AGB star known in the SMC, stressing the need for more such observations. Its photospheric chemistry reflects the growing chemical diversity among post-AGB stars and reinforces the complexity of AGB nucleosynthesis beyond current theoretical models. 
\end{abstract}

\section{Introduction}
\label{sec:intro}
The origin of heavy elements beyond iron ($A\,>\,60$) is primarily attributed to neutron-capture nucleosynthesis, with the slow neutron-capture process (\textit{s}-process) playing a dominant role in low- and intermediate-mass stars ($\sim$0.8–8~\Msun). The \sprocess operates under lower neutron densities ($n\,\lesssim10^{12}$ cm$^{-3}$) and over longer timescales, allowing unstable nuclei to undergo $\beta$-decay between successive neutron captures \citep[see][references therein]{iben74, Busso1999, Lugaro2023}. Asymptotic giant branch (AGB) stars are well established as the primary astrophysical site for the \sprocess \citep[e.g.][]{gallino98, Lugaro2003, Herwig2005, karakas14a}. The detection of technetium (Tc), a short-lived radioactive element produced by the \textit{s}-process, in the atmospheres of AGB stars provides compelling observational evidence of ongoing \sprocess nucleosynthesis during this phase \citep{Merrill1952, Werner2015, Shetye2020}. The \sprocess proceeds along the valley of $\beta$-stability and is typically categorised into three components based on neutron exposure: the weak component synthesises light \sprocess (ls) elements (e.g., Sr, Y, Zr), the main component synthesises heavier \sprocess (hs) elements (e.g., Ba, La, Nd); and the strong component is responsible for the heaviest stable \sprocess elements, including Pb and Bi \citep[e.g.][]{Lugaro2012, karakas14a}. The \hsls\ ratio is widely used as a diagnostic of neutron exposure and \sprocess efficiency during AGB nucleosynthesis. 

However, observationally constraining AGB nucleosynthesis remains challenging due to the extended, pulsating atmospheres, strong mass loss, molecular opacities, and circumstellar dust envelopes of AGB stars \citep{abia08, perez-mesa19}. Post-asymptotic giant branch (post-AGB) stars offer a valuable alternative. These luminous, short-lived objects have recently departed the AGB and retain surface compositions reflective of their prior nucleosynthetic histories \citep[see][]{vanwinckel03, Kamath2022, devika23}. Their relatively high photospheric temperatures (\Teff~$\sim$3000–30,000 K) allows access to a wide range of atomic lines in the optical regime, enabling detailed chemical abundance analyses of elements from CNO to heavy \sprocess elements well beyond the Ba peak \citep[e.g.,][]{Vanwinckel2000, vanaarle11, desmedt12, deSmedt2015, desmedt2016, Kamath2022}. These properties make post-AGB stars exquisite targets for constraining stellar evolution and nucleosynthesis models \citep{Kamath2022Universe}.

To date, $\sim$18 single \sprocess enriched post-AGB stars have been well studied in the Galaxy \citep[see][and references therein]{Kamath2022} $\sim$1 in the Small Magellanic Cloud (SMC) \citep[e.g.,][]{desmedt12} and $\sim$6 in the Large Magellanic Cloud (LMC) \citep[e.g.,][]{vanaarle11, desmedt12, deSmedt2015}. Analyses of these stars have uncovered significant discrepancies between observed surface abundances and those predicted by theoretical nucleosynthesis models. These discrepancies have sparked renewed interest in alternative neutron-capture processes. This has motivated the proposal of the intermediate neutron-capture process (\textit{i}-process), which operates at neutron densities between those of the \sprocess and \rprocess ($n~\sim10^{13}$–$10^{15}$ cm$^{-3}$) \citep[see][references therein]{Hampel2016, Choplin2021, Choplin2024}. These models show promising agreement for elements between the first and second \sprocess peaks. However, the success of such models often hinges on finely tuned parameters such as overshoot efficiencies or neutron exposures, limiting their general applicability. Notably, even in models that match the general abundance trends, significant discrepancies remain for lead (Pb), highlighting its role as a particularly sensitive diagnostic of neutron-capture processes—especially in metal-poor environments \citep{Kamathuniverse2021}.

Previous abundance studies of single \sprocess enriched post-AGB stars have been valuable but often piecemeal. A recent study by \citet{Kamath2022} analysed a larger, comprehensive Galactic sample of $\sim$30 single post-AGB stars, with luminosities and initial masses derived from Gaia DR3 parallaxes. This study exemplifies the remarkable chemical diversity among stars with similar stellar parameters and luminosities, revealing both \sprocess enriched and non-enriched stars, and challenging existing paradigms of AGB nucleosynthesis. Furthermore, the discovery of \sprocess enriched post-AGB binaries in both the Galaxy and the MCs further complicates the picture \citep{Menon2024}. Together, these findings point to a more nuanced and diverse range of neutron-capture signatures than previously appreciated.

The growing body of observational and theoretical work on post-AGB stars highlights that each star may reflect a distinct nucleosynthetic history \citep[see e.g.,][]{Kamath2022, devika23, Menon2024}. While large-sample studies reveal population-wide trends, detailed studies of individual stars remain essential for decoding the complex interplay of processes that shape surface abundances.

In this work, we present a detailed chemical abundance analysis of J003643.94$-$723722.1 (hereafter J003643), the most carbon-rich and one of the most \sprocess enriched post-AGB stars identified in the Small Magellanic Cloud (SMC). We compare the derived abundances with predictions from a range of theoretical stellar evolutionary models to investigate the underlying enrichment processes. J003643's well-constrained luminosity and initial mass, coupled with its unique chemical fingerprint, provide fresh insight into \sprocess nucleosynthesis in post-AGB stars. To better understand the chemical diversity observed in post-AGB stars, we compare some of the key abundance ratios of J003643 with a broader sample of post-AGB stars, including single \sprocess enriched, non-\sprocess enriched, and binary \sprocess enriched post-AGB stars. In addition, we aim to investigate Pb abundances in post-AGB stars using both local thermodynamic equilibrium (LTE) and a toy model to make a preliminary estimate of non-local thermodynamic equilibrium (NLTE) effects, thereby refining Pb abundance estimates and addressing longstanding discrepancies between observational data and theoretical nucleosynthesis predictions.

This paper is organised as follows: Section~\ref{sec:target_data} presents an overview of the target and describes the photometric and spectroscopic datasets. Section~\ref{sec:spectral_analysis} outlines the derivation of atmospheric parameters and chemical abundances. Section~\ref{sec:lum_sed} describes the luminosity derivation. Section~\ref{sec:abg_models} introduces the stellar evolutionary models used in this study. Section~\ref{sec:discussion} compares J003643 to the comparative sample of post-AGB stars and explores its nucleosynthetic origin by comparing with theoretical model predictions. We also assess the Pb discrepancy using both LTE and NLTE techniques. Section~\ref{sec:conclusion} summarises the main findings and implications of this study.

\section{Target Details}
\label{sec:target_data}
The primary target of this study, J003643, was first identified as a post-AGB candidate in the SMC through a comprehensive low-resolution spectroscopic survey by \citet{Kamath2014}. In the same study, J003643 was classified as a shell-source based on its characteristic double-peaked spectral energy distribution (SED; see Figure~\ref{fig:SED}). Shell-sources are typically thought to be single post-AGB stars that exhibit a distinct double-peaked SED, where the first peak in the near-IR corresponds to the stellar photosphere, while the second peak in the mid-infrared is attributed to thermal emission from a detached, expanding shell of cold circumstellar dust material ejected during the AGB phase \citep{vanwinckel03,Gezer2015}. 

\subsection{Photometric Data}
\label{sec:photometric_data}

\begin{figure}
        \includegraphics[width=\columnwidth]{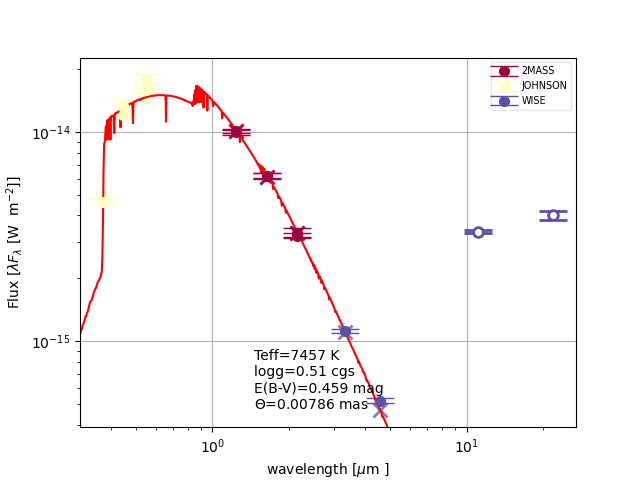}
    \caption{SED of J003643. The data points represent dereddened photometry, while the red line corresponds to the best-fitting scaled model atmosphere. See Section~\ref{sec:lum_sed} for details on the SED fitting.}
    \label{fig:SED}
\end{figure}

Photometric data for J003643 were retrieved from the Vizier database \citep{Vizier2000} to construct its SED (see Figure~\ref{fig:SED}). The optical and near-IR photometry characterises the photospheric emission of the post-AGB star, while mid- and far-IR photometry provides insight into the properties of the circumstellar environment.  

For the optical and near-IR regime, we utilised data from the UBVRI Johnson-Cousins photometric system \citep{Bessell1990}. Mid- and far-IR fluxes were obtained from the 2MASS All-Sky Catalog of Point Sources \citep{Cutri2003} and the WISE All-Sky Data Catalog \citep{Cutri2012}. The full set of photometric magnitudes for J003643 is presented in Table~\ref{tab:photometry}. A comprehensive list of commonly used photometric catalogues can be found in Appendix A of \citet{oomen18}.  

The resulting SED of J003643 is shown in Figure~\ref{fig:SED}. Details of the SED fitting procedure are described in Section~\ref{sec:lum_sed}.  

\input{Table1}

\subsection{Spectroscopic Observations and Data Reduction}
\label{sec:spectro_data}

High-resolution optical spectra of J003643 were obtained using the Ultraviolet and Visual Echelle Spectrograph (UVES) \citep{UVES}, mounted on the 8 m UT2 Kueyen Telescope of the Very Large Telescope (VLT) array at the European Southern Observatory’s (ESO) Paranal Observatory in Chile (Program ID: 092.D-0485). These observations were part of a broader effort to obtain high-quality spectroscopic data for post-AGB stars. To maximise wavelength coverage, the dichroic beam-splitter was employed, yielding spectra that span 3280–4560~\AA\ for the blue arm and 4726–5800~\AA\ and 5817–6810~\AA\ for the lower and upper regions of the red arm CCD chip, respectively. Due to the physical separation between the three UVES CCDs, small spectral gaps exist between 4560–4726~\AA\ and 5800–5817~\AA. Each spectral range was observed separately with a consistent exposure time. The resolving power of UVES for these observations ranges between $\sim$60,000 and $\sim$65,000.  

\begin{figure}
        \includegraphics[width=\columnwidth]{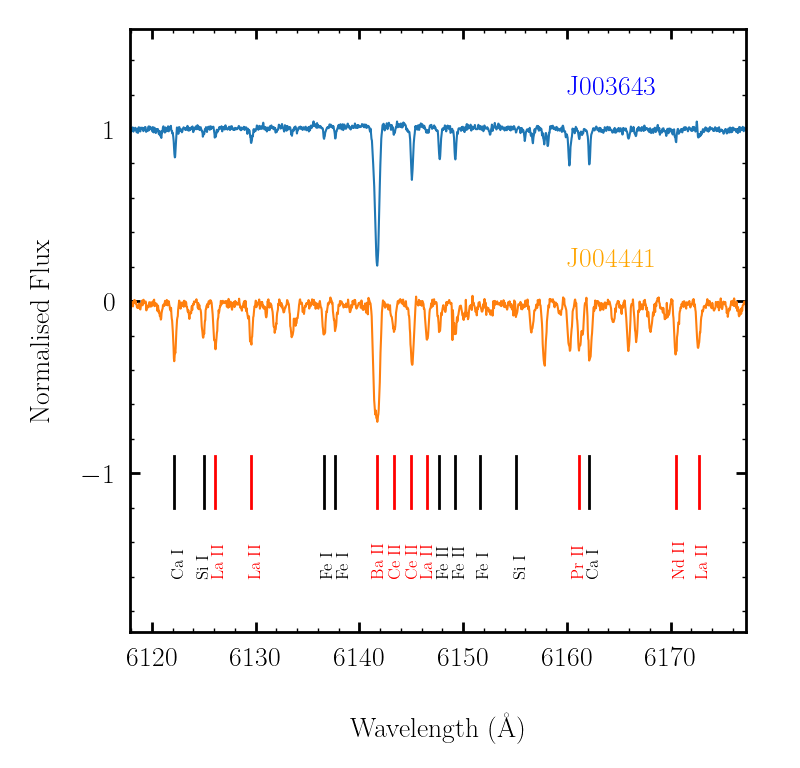}
    \caption{Normalised spectra of the target star J003643 (upper) and one of the most \sprocess enriched post-AGB stars, J004441 (lower), from \citet{desmedt12} for comparison. The spectra have been normalised and shifted to zero velocity for illustrative purposes. The red and black vertical lines mark the positions of selected \sprocess and non-\sprocess elements, respectively.}
    \label{fig:Spectra}
\end{figure}

Figure~\ref{fig:Spectra} presents a spectral region of the target star J003643 (upper) alongside J004441 (lower), one of the most \sprocess enriched post-AGB stars from \citet{desmedt12}, for comparison. This figure illustrates the quality of the UVES spectra and highlights the detection of specific \sprocess elements in J003643.

The observational details, including spectral coverage and signal-to-noise (S/N) ratios per observation, are provided in Table~\ref{tab:log}. The S/N ratio is lower at blue wavelengths due to increased atmospheric extinction, reddening effect and reduced detector sensitivity. Additionally, the echelle spectra exhibit non-uniform S/N owing to the blaze function, which results in a higher S/N near blaze wavelengths where the spectrograph operates most efficiently. The final weighted mean spectrum has an S/N of 30 in the blue region and 55 in the red region.
\input{Table2}

The pre-reduced UVES spectra of J003643 were retrieved from the ESO Archive Science Portal. The normalisation of the reduced spectrum was performed by fitting fifth-order polynomials within small spectral windows using interactively selected continuum points. This step is particularly crucial in the blue spectral region, where significant line crowding makes continuum placement challenging. The primary source of measurement uncertainty in abundance determinations for lines in this region arises from continuum placement errors.

The radial velocity of J003643 was determined by measuring the central wavelengths of several well-identified atomic lines using Gaussian profile fitting. The Doppler shift equation was then applied to derive a heliocentric radial velocity of v\,$= 133.9 \pm 0.26$ \kms, which is consistent with SMC membership, given the mean heliocentric velocity of the SMC being $\sim$160 \kms\ \citep{Richter1987}.

After normalisation and radial velocity correction of all the spectra (blue and red), a weighted mean merging was performed to obtain a single final spectrum, which was subsequently used for a detailed spectroscopic analysis of J003643. A significant portion of the blue spectrum (3280–4560~\AA) exhibits an S/N too low (<\,30) for reliable abundance determination. Consequently, these wavelength ranges were excluded from the spectroscopic analysis of J003643.

\section{Spectroscopic Analysis of J003643}
\label{sec:spectral_analysis}
The spectroscopic analysis of J003643, including the determination of accurate atmospheric parameters and the derivation of chemical abundances for all identifiable elements, was conducted using E-iSpec \citep{Mohorian2024}, a custom Python wrapper for iSpec \citep{Blanco2014a,Blanco2019}. E-iSpec interfaces with the LTE spectrum synthesis tool MOOG \citep{sneden1973}. A detailed description of E-iSpec and the spectroscopic analysis methodology is provided in \citet{Mohorian2024}. Our approach closely follows the methods outlined in \citet{Menon2024} and \citet{Mohorian2024}.  

The atmospheric parameters and elemental abundances were determined using carefully selected spectral lines that are isolated, unblended, and non-saturated. To minimise spectral noise, we included only lines with equivalent widths (EW) greater than 5 m\AA, while excluding those exceeding 150 m\AA\ to avoid saturation effects.  

A summary of the atmospheric parameter determination and chemical abundance derivation is presented in Sections~\ref{sec:atmos_param} and \ref{sec:abund_analysis}, respectively.  

\subsection{Atmospheric Parameter Determination}
\label{sec:atmos_param}
The atmospheric parameters of J003643 were determined using Fe I and Fe II lines, following the approach described in \citet{Mohorian2024} and \citet{Menon2024}. The EWs for each line were computed iteratively by comparing the theoretical EWs with the observed values.  

The effective temperature (\Teff) was derived through excitation equilibrium, ensuring that Fe I abundances remain independent of excitation potential (EP). The surface gravity (\logg) was determined via ionisation equilibrium, enforcing consistency between Fe I and Fe II abundances. The microturbulent velocity (\mv) was obtained by requiring that the derived Fe abundances are independent of the reduced equivalent width (RW) of individual lines. Uncertainties in the atmospheric parameters were estimated using the covariance matrix from the non-linear least-squares fitting algorithm in E-iSpec.  
\input{Table3}

The results of the atmospheric parameter analysis for J003643 are presented in Table~\ref{tab:atmosParam_J003643} and discussed in Section~\ref{sec:results}. The values in the last column of Table~\ref{tab:atmosParam_J003643} correspond to the atmospheric parameters derived from low-resolution AAOmega spectra by \citet{Kamath2014}.

\subsection{Derivation of Chemical Abundances}
\label{sec:abund_analysis}
The chemical abundances of different elements of J003643 were derived using both the EW method and spectral synthesis fitting (SSF), both implemented in E-iSpec, following a similar approach to \citet{Menon2024}. In the EW method, the observed EWs were compared with theoretical EWs based on synthetic spectra from a model atmosphere (ATLAS9; \citealt{castelli03}). The assumed abundances were iteratively adjusted until the observed and predicted EWs converged, yielding the best-fit values.

\begin{figure*}
    \centering
    \begin{subfigure}[b]{0.49\linewidth}
        \centering
        \includegraphics[width=\linewidth]{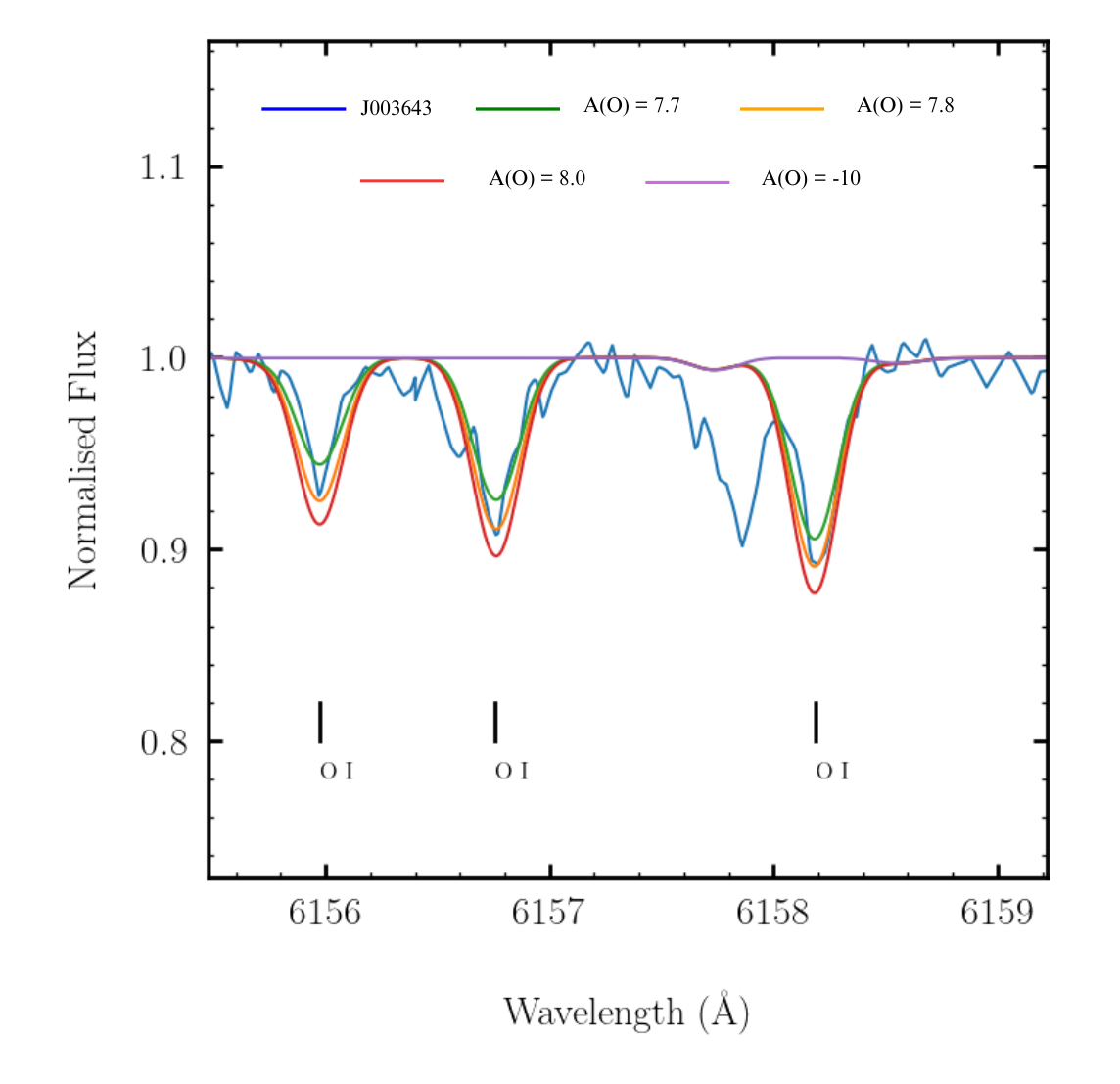}
        \caption{SSF of the O I lines at 6155.971~\AA, 6156.755~\AA, and 6158.187~\AA\ for J003643.}
        \label{fig:O}
    \end{subfigure}
    \hfill
    \begin{subfigure}[b]{0.49\linewidth}
        \centering
        \includegraphics[width=\linewidth]{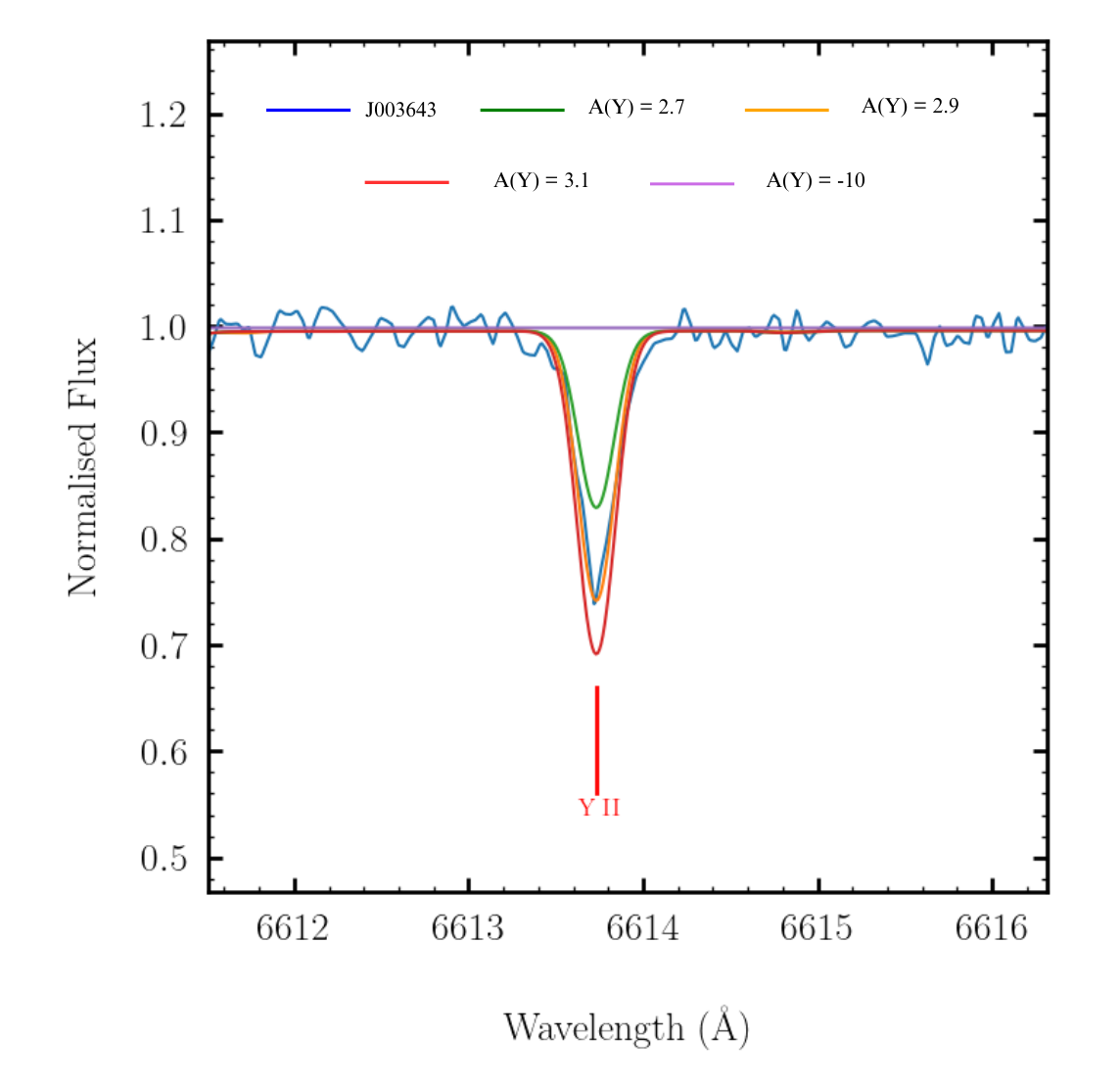}
        \caption{SSF of the Y II line at 6613.731~\AA\ for J003643.}
        \label{fig:Y}
    \end{subfigure}
    \hfill 
    \begin{subfigure}[b]{0.49\linewidth}
        \centering
        \includegraphics[width=\linewidth]{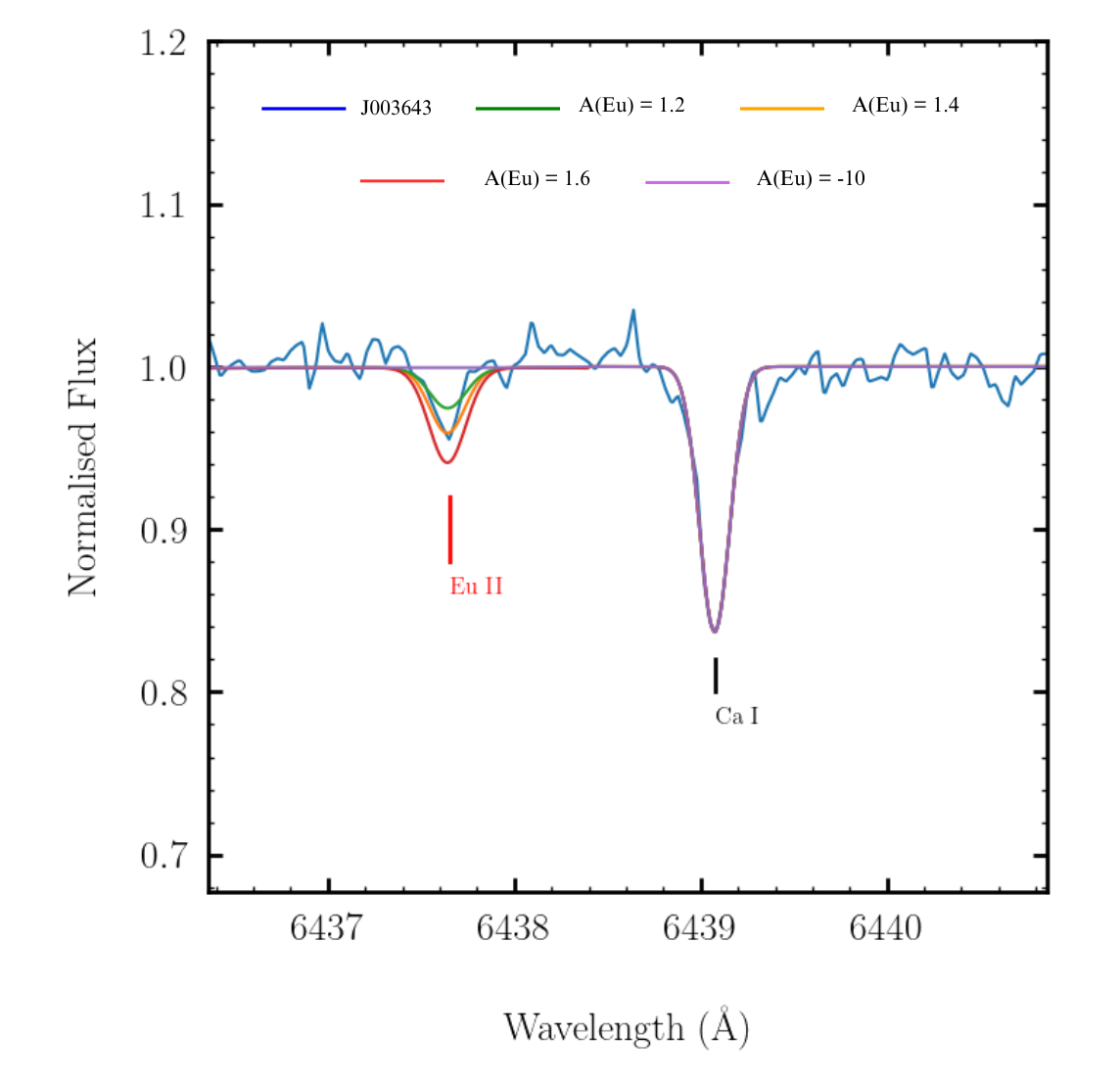}
        \caption{SSF of the Eu II line at 6437.640~\AA\ for J003643. }
        \label{fig:Eu}
    \end{subfigure}
    \begin{subfigure}[b]{0.49\linewidth}
        \centering
        \includegraphics[width=\linewidth]{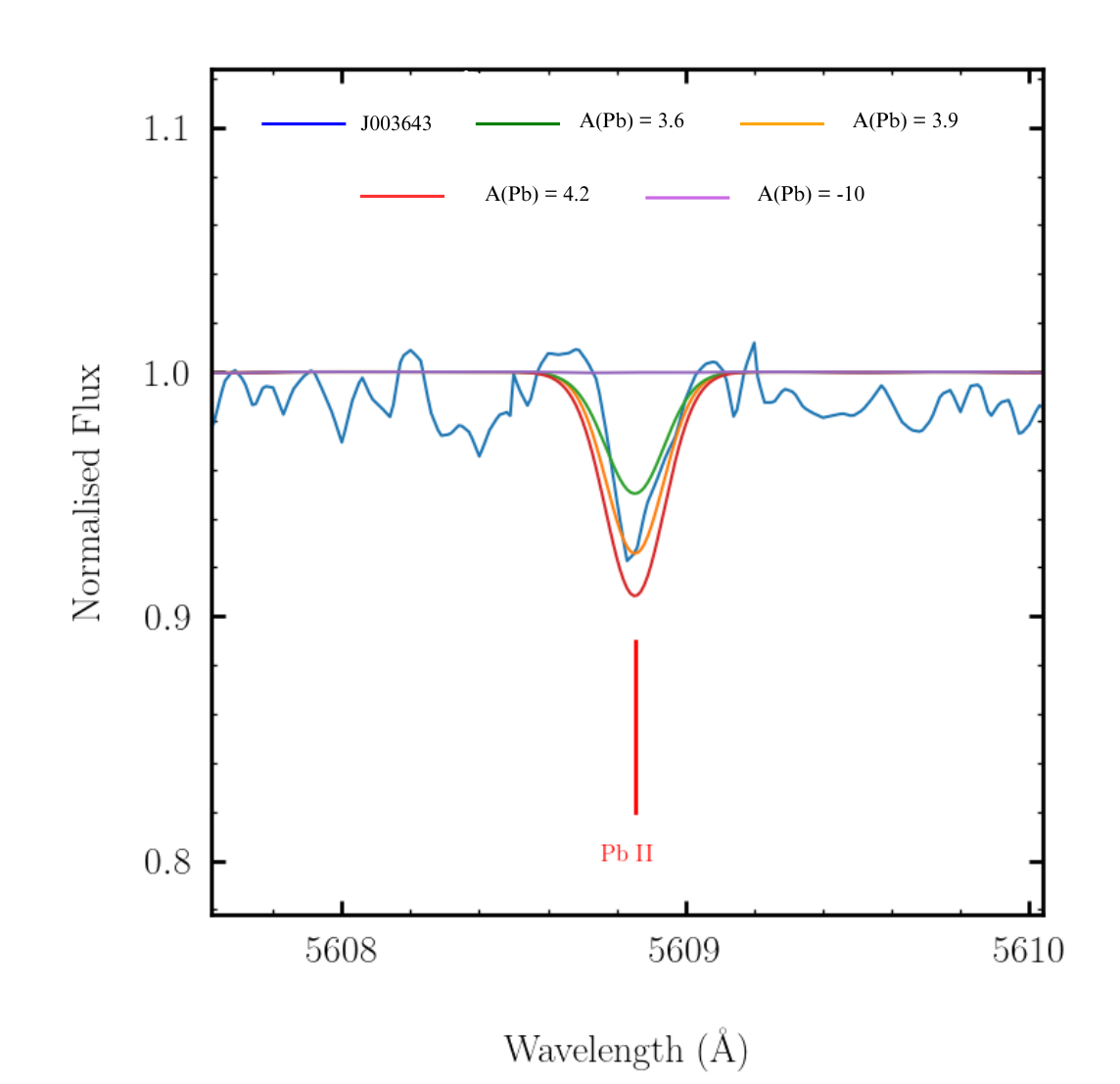}
        \caption{SSF of the Pb II line at 5608.853~\AA\ for J003643.}
        \label{fig:Pb}
    \end{subfigure}
    \caption{SSF of the O I lines at 6155.971~\AA, 6156.755~\AA, and 6158.187~\AA\ (top left), Y II line at 6613.731~\AA\ (top right), Eu II line at 6437.640~\AA\ (bottom left), and Pb II line at 5608.853~\AA\ (bottom right) for J003643. Here, A(X) denotes the logarithmic absolute abundance $\log\epsilon(X)$ of the respective element, and A(X)\,$=\,-10$ indicates a synthetic spectrum computed with zero abundance of that element.}
    \label{fig:synth_Spec}
\end{figure*}

The SSF method involved comparing the observed spectra with synthetic spectra generated from model atmospheres, using a chi-square fitting procedure to iteratively refine the model parameters and determine precise chemical abundances. Instrumental and macroturbulent broadening were incorporated through Gaussian convolution, consistent with the UVES resolving power (R$\,\approx60,000$) and with macroturbulent velocities expected in cool, low-gravity stellar atmospheres. Examples of SSF applied to the spectral lines of four elements are shown in Figure~\ref{fig:synth_Spec}: O I at 6155.971~\AA, 6156.755~\AA, and 6158.187~\AA, Y II at 6613.731~\AA, Eu II at 6437.640~\AA\ and Pb II at 5608.853~\AA. These illustrate the sensitivity of the derived abundances to both line strength and local continuum placement. SSF was performed alongside the EW method for these elements to ensure consistency and to validate the robustness of the derived abundances.  

The lines used for abundance derivation were selected based on the line selection criteria outlined in Section~\ref{sec:spectral_analysis}. The linelist used for the spectroscopic analysis of J003643 is provided in Table~\ref{tab:linelist_J003643}.

The uncertainties in the derived abundances were estimated using the method outlined in \citet{Deero2005}. E-iSpec calculates the errors as standard deviations (\sigmalinetoline) for each line, while the total error (\sigmatotal) is computed as the quadratic sum of \sigmalinetoline, uncertainties due to atmospheric parameters, and the \feh\ error. To assess the sensitivity of the abundances to stellar parameters, we recomputed the abundances by varying \Teff, \logg, and \mv within their respective uncertainties (see Table~\ref{tab:atmosParam_J003643}). A \sigmalinetoline\ of 0.20 dex was adopted for species with only a single measurable line, reflecting the expected line-to-line scatter.
\input{Table4}

The derived chemical abundances for J003643 are presented in Table~\ref{tab:abund_J003643}, and the result of the analysis is discussed in Section~\ref{sec:results}. The elements in Table~\ref{tab:abund_J003643} are arranged by atomic number (Z), with the corresponding solar abundances (\logepsilonsun) taken from \citet{Asplund2009}. The table also includes the number of identified lines per element (N), the element-to-iron ratio (\xfe), the total uncertainty on \xfe\ (\sigmatotal), the absolute abundance (\logepsilon), and the line-to-line scatter (\sigmalinetoline). Despite the inability to determine the abundances of several key nucleosynthesis elements, Table~\ref{tab:abund_J003643} presents quantified abundances for a wide range of \sprocess elements.

Notably, this study presents the first direct measurement of the Pb abundance in a post-AGB star, derived from the Pb II line at 5608.853~\AA, which was the only detectable Pb line in J003643 (further details on Pb are provided in Section~\ref{sec:lead}). In contrast, previous studies of post-AGB stars have only been able to derive upper limits on Pb abundance using SSF \citep[e.g.,][]{desmedt2016}, primarily due to the intrinsic weakness of Pb lines and line blending effects, which have hindered precise abundance measurements.  

\subsubsection{NLTE Effects}
\label{sec:nlte}
Departures from LTE, if not accounted for, can introduce inaccuracies in spectroscopic analyses and lead to biased conclusions. To assess their impact, we tested possible NLTE effects on spectral lines of C I, O I, Mg I, and Ca I. Weak lines of Ti II \citep{2024A&A...687A...5M} and Fe II \citep{lind2012NLTE} are expected to show relatively mild departures from LTE. Therefore, we adopted \feh\ derived from Fe II lines as the metallicity and applied NLTE corrections to the elemental abundances of C, O, Mg, and Ca accordingly. The results, including uncertainties, are presented in Table~\ref{tab:abund_J003643} under the column "\xfenlte". To calculate the uncertainties of the NLTE abundances, we adopt the same systematic error as in the LTE analysis, the component that quantifies the sensitivity of the abundances to uncertainties in \Teff, \logg, and \mv\ by recomputing them within their quoted errors, since the NLTE corrections themselves are essentially invariant to these parameter shifts. We then compute the \sigmalinetoline\ of the NLTE abundances (eNLTE), which is found to be very similar to that in LTE. Consequently, the total uncertainty \sigmatotal\ for the NLTE abundances matches the LTE value to two decimal places.

The NLTE calculations were performed using \texttt{Balder} \citep{amarsi2018Balder}, which is based on \texttt{Multi3D} \citep{Multi3D2009}. These calculations were carried out using the same ATLAS model atmosphere of J003643 employed for LTE abundance determinations in this study. The abundances were computed for five values in steps of 0.5 dex around the 1D LTE value. The model atoms were adopted from \citet{2019A&A...624A.111A} for C I, \citet{2018A&A...616A..89A} for O I, and \citet{2021A&A...653A.141A} for Mg I and Ca I.  

A caveat of this test is that the adopted model atoms were originally calibrated for stars with lower effective temperatures and higher atmospheric densities than J003643. Nevertheless, the results provide insight into the extent to which departures from LTE might affect our abundance determinations.

\subsection{Results of Spectroscopic Analysis: Atmospheric Parameters and Chemical Abundances}
\label{sec:results}
In this section, we present the results of the spectroscopic analysis of J003643.

The atmospheric parameters derived for J003643 are summarised in Table~\ref{tab:atmosParam_J003643}. Our analysis confirms that J003643 is an A-type post-AGB star with an effective temperature of \Teff\,= $7552\pm91$ K. It has a low surface gravity (\logg\,= $1.04\pm0.16$ dex) and an iron abundance of \feh\,= $-0.97\pm0.11$ dex. Although the iron abundance of J003643 is low, it is comparable to the mean metallicity of the SMC ($\feh\,= -0.7$ dex) \citep{luck1998}.

\begin{figure}
    \includegraphics[width=\columnwidth]{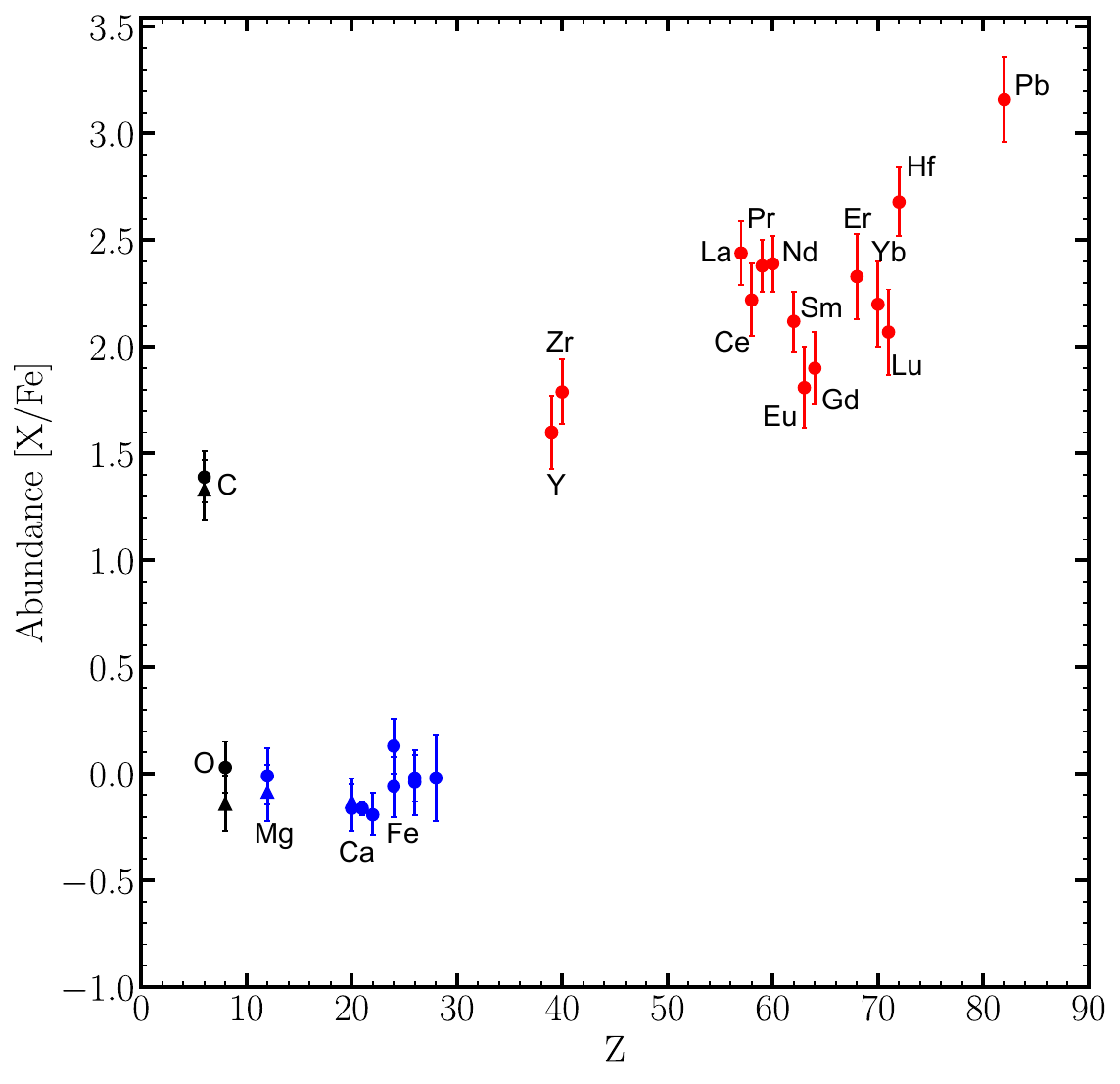}
    \caption{Spectroscopically derived abundances of J003643. The error bars represent the total uncertainties (\sigmatotal). The data points are colour-coded: black for CNO elements, blue for Fe-peak elements, and red for \sprocess elements. Circles represent LTE abundances, while triangles represent NLTE abundances (for details on NLTE, see Section~\ref{sec:nlte}). Some elements are labelled for reference.}
    \label{fig:abund_J003643}
\end{figure}

The derived chemical abundances for J003643 are presented in Table~\ref{tab:abund_J003643}. Figure~\ref{fig:abund_J003643} illustrates the abundance distribution (\xfe) as a function of atomic number (Z).
In Figure~\ref{fig:abund_J003643}, the circles represent LTE abundances, while the triangles correspond to NLTE abundances (for details on NLTE, see Section~\ref{sec:nlte}). The NLTE abundances, where available, generally agree with the LTE results within the uncertainties. 

Figure~\ref{fig:abund_J003643} shows that J003643 has a \cfe\ of 1.33$\pm$0.14 dex and an \ofe\ of -0.08$\pm$0.20 dex. Notably, it has a C/O ratio of $12.58\pm0.25$ based on LTE abundances and $16.21\pm0.22$ from NLTE-corrected abundances. This is relatively higher than the typical C/O ratios of \sprocess enriched post-AGB stars ($\approx1$ to 3), primarily driven by J003643’s low oxygen abundance, relative to typical post-AGB stars (see Section~\ref{sec:full_sample} for details).

Additionally, Figure~\ref{fig:abund_J003643} also shows that J003634 is enriched in \sprocess elements. The \sprocess distributions and overabundances of LIM stars are commonly described using four observational indices: \lsfe, \hsfe, \sfe, and \hsls. To maintain consistency, we use the same elemental abundances as in \citet{desmedt2016} for calculating these indices. We define the hs index as the average of La, Ce, Nd, and Sm; the ls index as the average of Y and Zr; and the overall \sfe index as the mean of all six \sprocess elements. For J003643, we derive \lsfe\,= 1.72 $\pm$ 0.17 dex, \hsfe\,= 2.19 $\pm$ 0.11 dex, \hsls\,= 0.48 $\pm$ 0.12 dex and \sfe\,= $2.09\pm0.20$ dex.

Furthermore, while previous studies of \sprocess enriched post-AGB stars have only reported upper limits for Pb abundances based on SSF \citep[e.g.][]{desmedt2016}, this study provides the first direct Pb abundance measurement in a post-AGB star using both EW and SSF methods. The clearly detected Pb II line at 5608.853~\AA\ yields a consistent Pb abundance of \pbfe\,= 3.16 dex from EW fitting and \pbfe\,= 3.20 dex from SSF (see Figure~\ref{fig:Pb}), enabling a more robust comparison with theoretical nucleosynthesis predictions.

\section{Luminosity Determination Through SED Fitting}
\label{sec:lum_sed}
Accurately determining the luminosity of a post-AGB star and combining it with its derived atmospheric parameters and elemental abundances provides valuable insights into its evolutionary stage and facilitates the estimation of its initial mass. However, deriving the luminosities of post-AGB stars remains challenging due to several factors, including the effects of circumstellar and interstellar reddening on observational photometry \citep[see e.g.,][]{Kamath2022, devika23}.  

To determine the luminosity of J003643, we utilised its SED fitting parameters, following the approach outlined in \citet{Kluska2022} and \citet{Kamath2022}. In summary, the total line-of-sight reddening or extinction parameter ($E(B-V)$) was determined by minimising the difference between the observed optical fluxes and the reddened photospheric models. This total reddening accounts for both interstellar and circumstellar extinction. We assumed that the total reddening follows the wavelength dependency of the interstellar extinction law \citep{cardelli89} with $R_V\,= 3.1$. While the extinction law in the circumstellar environment likely differs from the interstellar law, investigating this aspect is beyond the scope of this study.  

For the atmospheric models, we adopted the Kurucz models \citep{castelli03} with parameters derived from the spectroscopic analysis in Section~\ref{sec:atmos_param} (see Table~\ref{tab:atmosParam_J003643}). The SED fitting was performed by interpolating within the $\chi^2$ landscape, using models centred on the spectroscopically determined parameters and applying variations of $\Delta$\Teff\ $\pm 500$ K, $\Delta$\logg\ $\pm 0.5$ dex, and $\Delta$\feh\ $\pm 0.5$ dex (see Section~\ref{sec:atmos_param}). The resulting SED of J003643 is shown in Figure~\ref{fig:SED}.
\input{Table6}

The bolometric luminosity or SED luminosity (\Lsed) of J003643 was derived by integrating the flux under the dereddened photospheric SED model, assuming an average distance of $62.1 \pm 1.9$ kpc to the SMC \citep{graczyk}. The derived SED luminosity (\Lsed) and its corresponding upper and lower limits for J003643 are $7623^{8606}_{5761}~\Lsun$. To estimate the uncertainties in luminosity, we incorporated the upper and lower limits of the reddening values, as outlined in Table~\ref{tab:summary}. 

\section{Stellar Evolutionary Models: ATON, MONASH and FRUITY}
\label{sec:abg_models}
Comparing the derived chemical abundances of a star with stellar evolutionary models is key to understanding its evolution and nucleosynthesis. In this section, we compare the spectroscopically derived abundances of J003643 with predictions from three independent stellar evolutionary models: ATON (Section~\ref{sec:aton}), MONASH (Section~\ref{sec:monash_models}) and FRUITY (Section~\ref{sec:fruity_models}), to assess nucleosynthetic efficiency and constrain parameters such as initial mass and mixing.

\subsection{ATON}
\label{sec:aton}
To facilitate meaningful comparison with theoretical stellar evolutionary models, we first estimated the initial mass of J003643, a key parameter that governs stellar evolution and chemical yields. For this, we utilised evolutionary tracks computed with the \texttt{ATON} stellar evolution code \citep{ventura98}.

\texttt{ATON} stellar evolution code models the structure and nucleosynthesis of LIM stars, employing the Full Spectrum of Turbulence (FST) model for convective mixing \citep{ventura98, Ventura2008, ventura18}. It includes treatments of mass loss, convective boundary mixing, and thermal pulses during the AGB phase. The code tracks the evolution of light-element surface abundances, particularly C, N, and O, shaped by internal mixing processes. Although it does not compute detailed heavy-element nucleosynthesis, ATON provides robust predictions for CNO enrichment and structural evolution relevant to AGB modelling \citep[see e.g.,][]{devika23}.

In this study, we adopt an initial metallicity of Z = 0.0014, and use \texttt{ATON} to model the stellar evolution of J003643 based on its spectroscopically derived luminosity, atmospheric parameters, and surface carbon and oxygen abundances, following the methodology of \citet{devika23}.

\begin{figure}
    \includegraphics[width=\columnwidth]{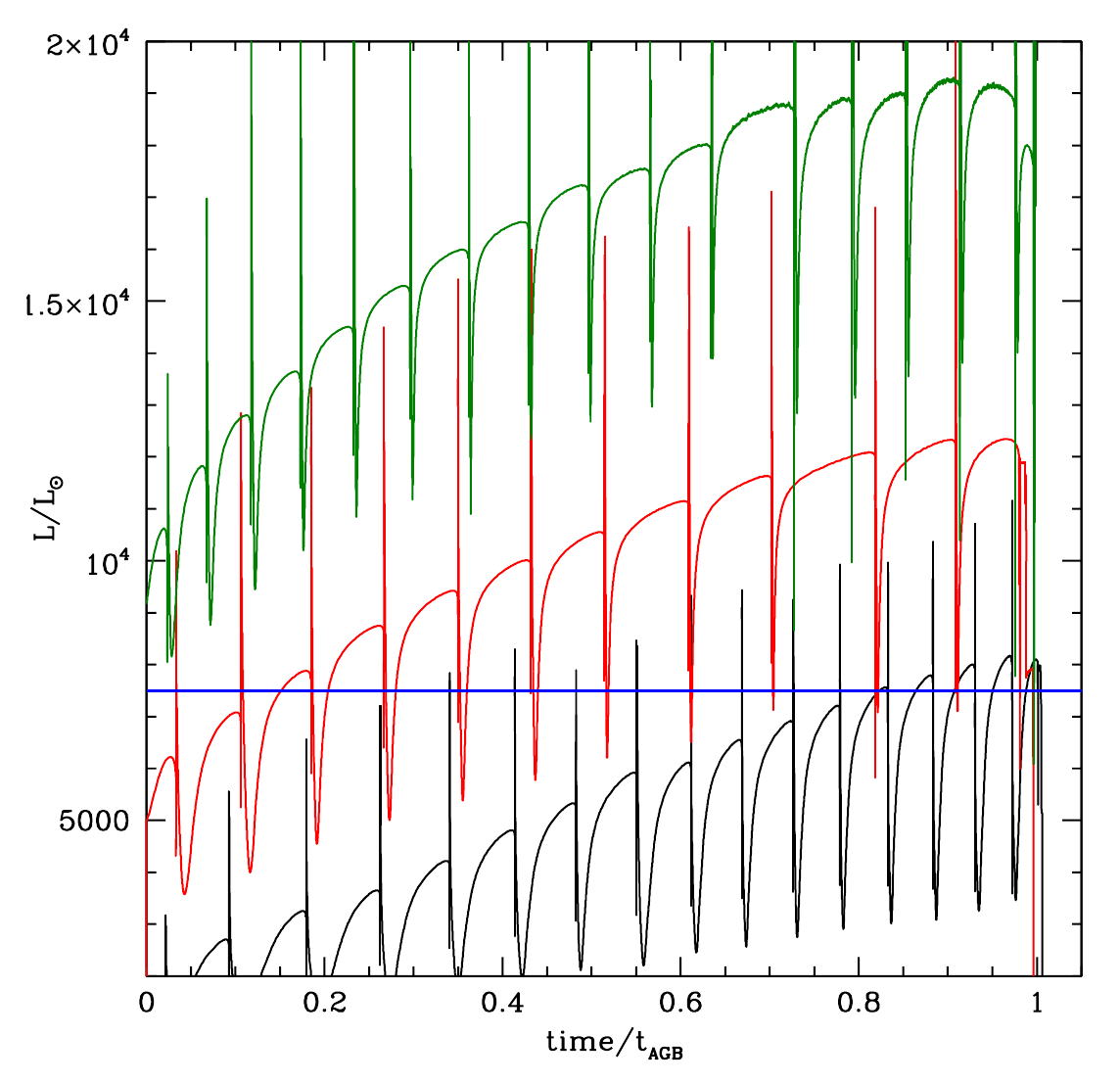}
    \caption{Luminosity evolution as a function of AGB lifetime for stars with an initial metallicity of Z = 0.0014. The models correspond to initial masses of $2~\Msun$ (black), $2.5~\Msun$ (red), and $3~\Msun$ (green). The blue horizontal line represents the derived SED luminosity of J003643 (see Section~\ref{sec:lum_sed}). See text for more details.}
    \label{fig:ATON1}
\end{figure}

\begin{figure}
    \includegraphics[width=\columnwidth]{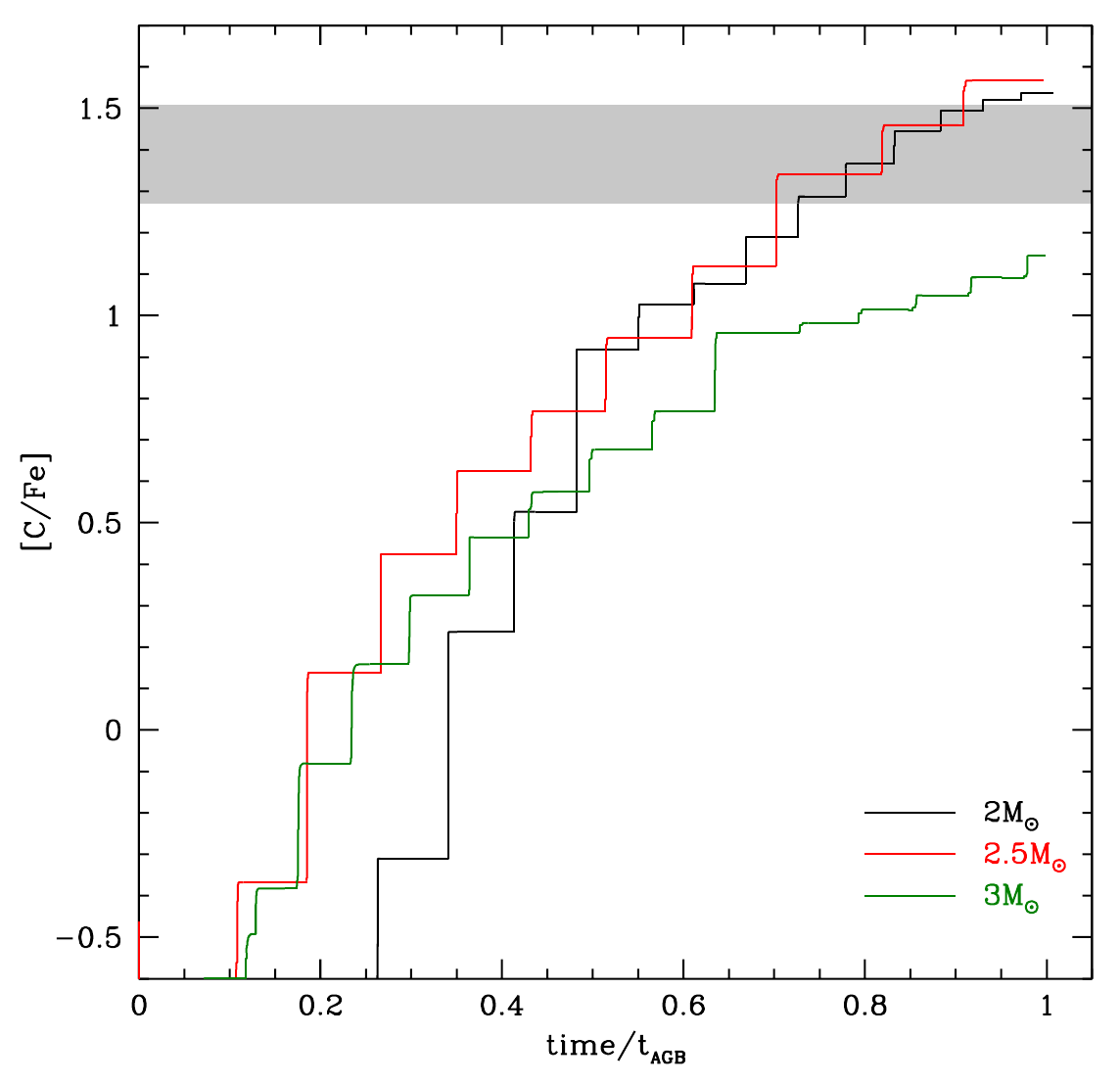}
    \caption{Surface \cfe\ abundance evolution as a function of AGB lifetime for stars with an initial metallicity of Z = 0.0014. The models correspond to initial masses of $2~\Msun$ (black), $2.5~\Msun$ (red), and $3~\Msun$ (green). The grey-shaded region indicates the spectroscopically derived carbon abundance of J003643 with its uncertainty (see Section~\ref{sec:abund_analysis}). See text for more details.}
    \label{fig:ATON2}
\end{figure}

Figures~\ref{fig:ATON1} and~\ref{fig:ATON2} illustrate the luminosity evolution and surface \cfe\ abundance variation as functions of the AGB lifetime (normalised to the total duration of the AGB phase) for stars with initial masses of $2~\Msun$ (black), $2.5~\Msun$ (red), and $3~\Msun$ (green). In Figure~\ref{fig:ATON1}, the blue horizontal line represents the SED-derived luminosity of J003643 (see Section~\ref{sec:lum_sed}), while in Figure~\ref{fig:ATON2}, the grey-shaded region indicates the spectroscopically derived carbon abundance of J003643 with its uncertainty (see Section~\ref{sec:abund_analysis}). We use \cfe\ rather than \ofe\ to constrain AGB evolution because carbon is gradually enhanced through third dredge-up (TDU) episodes, making it highly sensitive to the progenitor mass and internal nucleosynthesis. In contrast, oxygen shows little surface variation during the AGB phase for low-mass stars (<3~\Msun) and primarily reflects the composition of the progenitor gas. This distinction is confirmed by \citet{devika23}, demonstrating that \cfe\ correlates more tightly with stellar mass and TDU efficiency, while \ofe\ remains largely unaffected in the mass range relevant to post-AGB stars. Nevertheless, the \ofe\ predicted by the 2~\Msun\ model assuming no $\alpha$-enhancement is $\sim$0.2 dex, which is slightly higher but still consistent within uncertainties with the spectroscopically derived oxygen abundance of J003643 (\ofe\,= $-0.08 \pm 0.20$ dex).

The evolutionary tracks show that all three models cross evolutionary phases consistent with the estimated luminosity of J003643. Specifically, in the $2~\Msun$ model, the observed luminosity is reached toward the end of the AGB phase. In the $2.5~\Msun$ model, it occurs in the first half of the AGB lifetime and in the $3~\Msun$ model, the luminosity is attained at the beginning of the AGB phase. However, within the ATON grid, higher masses ($\geq3~\Msun$) can be ruled out, as hot-bottom burning (HBB) would prevent carbon enrichment, which contradicts the observed high \cfe\ in J003643. Similarly, within the ATON grid, lower masses (<2~\Msun) are also excluded, as they fail to reach the derived luminosity of J003643 (indicated by the blue line in Figure~\ref{fig:ATON1}) \citep{flavia15}.  

Furthermore, examining the surface carbon abundance evolution in Figure~\ref{fig:ATON2}, we find that the observed carbon abundance of J003643 aligns best with the final phases of the AGB. This suggests that J003643 most likely originated from a star with an initial mass of about $2~\Msun$, corresponding to a formation epoch of approximately 1 billion years ago. We note that the initial mass estimated here refers to the mass at the ZAMS, before the star evolved into its post-AGB phase.  

\subsection{MONASH}
\label{sec:monash_models}
The MONASH stellar evolution code \citep{Lattanzio1986, Karakas2014} was used to construct detailed stellar evolution models of J003643. The stellar evolutionary sequences allow us to calculate the detailed nucleosynthesis, including \sprocess elements, using a post-processing nucleosynthesis code. We refer to \citet{karakas16} for further details. For this study, we used the same version of the code and input physics described in \citet{Karakas2018}. The main uncertainties that we examined include the amount of TDU, which we can change by including convective overshoot at the base of the convective envelope \citep[e.g, see][]{kamath12}, and the size of the partially mixed zone (PMZ) in the post-processing calculations. In our post-processing calculations, the \textsuperscript{13}C pocket is formed by inserting protons into the top layers of the He-intershell during the deepest extent of each TDU event. The extent of the protons in the He-shell is called the PMZ. Observations of barium stars show a range of \sprocess abundances which likely indicate some natural variation in the size of \textsuperscript{13}C pockets \citep[e.g.,][]{Cseh2018}. For this reason, we experiment with calculating models with different size PMZ (we refer to \citet{Butain2017} for a detailed study of \textsuperscript{13}C pockets in our models, including a comparison to other studies). 

We experiment with models with initial masses between 1.5 to 2~\Msun\ with Z = 0.001 and find that the best match to the luminosity of J003643 near the end of the TP-AGB phase occurs in a 1.7~\Msun\ model. This is similar to the result from the ATON code (2~\Msun, see Section~\ref{sec:aton}), given that these are two independent stellar evolution codes and there is a slight difference in metallicity. We note that a 0.3~\Msun\ offset is well within the typical systematic uncertainties of TP‑AGB modelling. For the nucleosynthesis predictions (Figure~\ref{fig:MONASH}, green line), we adopt a PMZ mass of 0.006~\Msun, which provides the best overall match to the observed \sprocess enrichment. 

The results and the interpretation of our comparison between the spectroscopically derived abundances of J003643 and the MONASH models, including the impact of PMZ on \sprocess abundances, are discussed in Section~\ref{sec:model_disc}.

\subsection{FRUITY}
\label{sec:fruity_models}
To explore the impact of differences in input physics and numerical treatments across stellar evolution codes, we also utilised the stellar evolution and nucleosynthesis framework: FRUITY (Full-Network Repository of Updated Isotopic Tables and Yields) \citep{cristallo11}. FRUITY is designed to simulate the evolution and chemical enrichment of LIM stars, with a particular focus on \sprocess nucleosynthesis. These models span a range of initial masses (1.3–6.0~\Msun) and metallicities (Z = 0.00002 to Z = 0.03), making them well-suited for comparisons with stars observed in different galactic environments.

For this study, we use the \sprocess distribution from the latest FRUITY models, which incorporate the effects of mixing triggered by magnetic fields \citep{Vescovi2020}, to compare with the spectroscopically derived abundances of J003643. These updated FRUITY models incorporate a magnetic mixing mechanism as a new driver for the formation of the \textsuperscript{13}C pocket, the primary neutron source for \sprocess nucleosynthesis. Unlike previous versions \citep{cristallo11,Piersanti2013,Cristallo2015}, where the \textsuperscript{13}C pocket was formed via convective overshoot, this new approach accounts for buoyancy-driven magnetic instabilities, which naturally induce mixing within the He-intershell. This results in a self-consistent \textsuperscript{13}C pocket formation, influencing both neutron exposure and the overall efficiency of the \textit{s}-process.

In addition to this revised mixing mechanism, several other updates have been implemented in the latest FRUITY models, including the equation of state (EOS) and the nuclear reaction network \citep[see][for more details]{Vescovi2021}. For the models presented in this work, an updated initial chemical composition from \citet{Lodders2021} and revised values for C, N, O, and Ne from \citet{Magg2022} were used. Additionally, mass-loss prescriptions were updated using a revised period-mass-loss relation calibrated to the latest Galactic O- and C-rich giant data \citep{Ofner2018}, with pulsation periods derived from the period-mass-radius relation of \citet{Trabucchi2022}. The updated mass-loss rate is less efficient than that previously adopted in \citet{Vescovi2020, Vescovi2021}, further enhancing predictions of carbon and heavy-element abundances.

We tested models with initial masses between 1.5 and 2~\Msun\ at Z = 0.001 and found that the 2~\Msun\ magnetic-mixing FRUITY model provides the best match to the observed heavy-element abundances (see Figure~\ref{fig:fruity}, black line). We also explore the magnetic-mixing FRUITY model with an initially \rprocess enhanced composition, adopting a solar \rprocess distribution prescribed by \citet{Prantzos2020} scaled by +2 dex ([r/Fe]\,= +2 dex) from Ga to Bi (see Figure~\ref{fig:fruity}, magenta line).

The results and the interpretation of our comparison between the spectroscopically derived abundances of J003643 and the magnetic mixing FRUITY models are discussed in Section~\ref{sec:model_disc}.

\section{Discussion}
\label{sec:discussion}
In this section, we place the photospheric chemical composition of J003643 in a broader astrophysical context. We first compare its CNO and \sprocess abundance ratios with a relevant and comprehensive sample of post-AGB stars to investigate chemical trends and diversity (Section~\ref{sec:full_sample}). We then explore the origin of its chemical enrichment by comparing its abundance pattern against predictions from various stellar evolutionary models (Section~\ref{sec:model_disc}). Finally, we assess the longstanding Pb abundance discrepancy, including a first exploration of possible NLTE effects on Pb (Section~\ref{sec:lead}). 

\subsection{Comparative Analysis of Key Abundance Ratios in Post-AGB Stars}
\label{sec:full_sample}
The detailed spectroscopic analysis of J003643 revealed a relatively high C/O ratio (16.21) compared to other post-AGB stars ($\sim$1-3). To investigate this anomaly, we compared some of the key abundance ratios of J003643 (C/O, \cfe, \ofe, \sfe, \hsls) with a comprehensive sample of post-AGB stars, including single \sprocess enriched stars, single non-\sprocess enriched stars, and binary \sprocess enriched stars, spanning different galactic environments, the LMC, SMC and Galaxy. The full post-AGB sample (referred to hereafter as the "comparative sample") considered in this study, including their references, is summarised in Table~\ref{tab:full sample}.
\begin{figure*}
    \centering
    \begin{subfigure}[b]{0.49\linewidth}
        \centering
        \includegraphics[width=\linewidth]{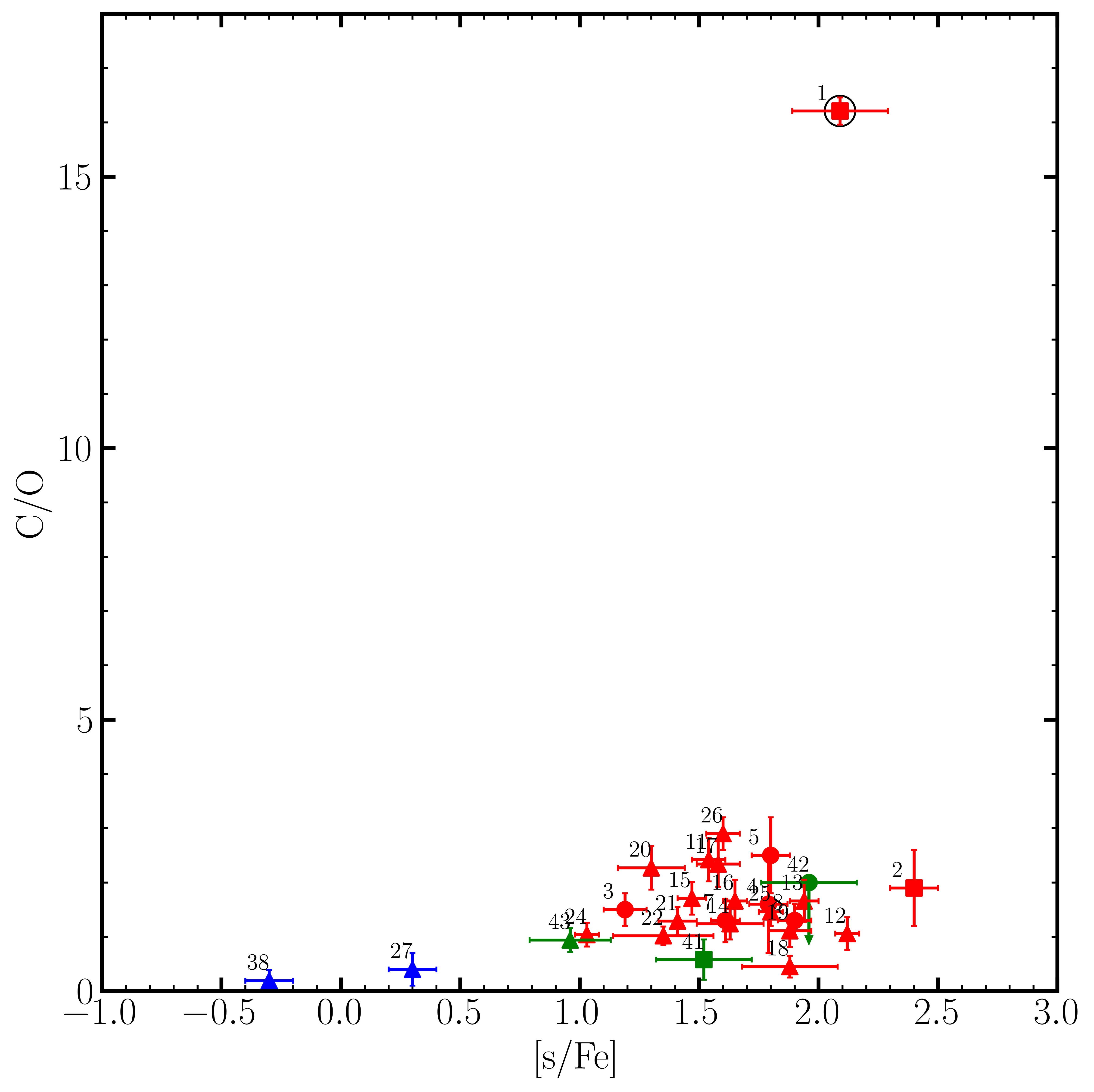}
        \caption{C/O ratio as a function of \sfe}
        \label{fig:CO_sFe}
    \end{subfigure}
    \hfill
    \begin{subfigure}[b]{0.49\linewidth}
        \centering
        \includegraphics[width=\linewidth]{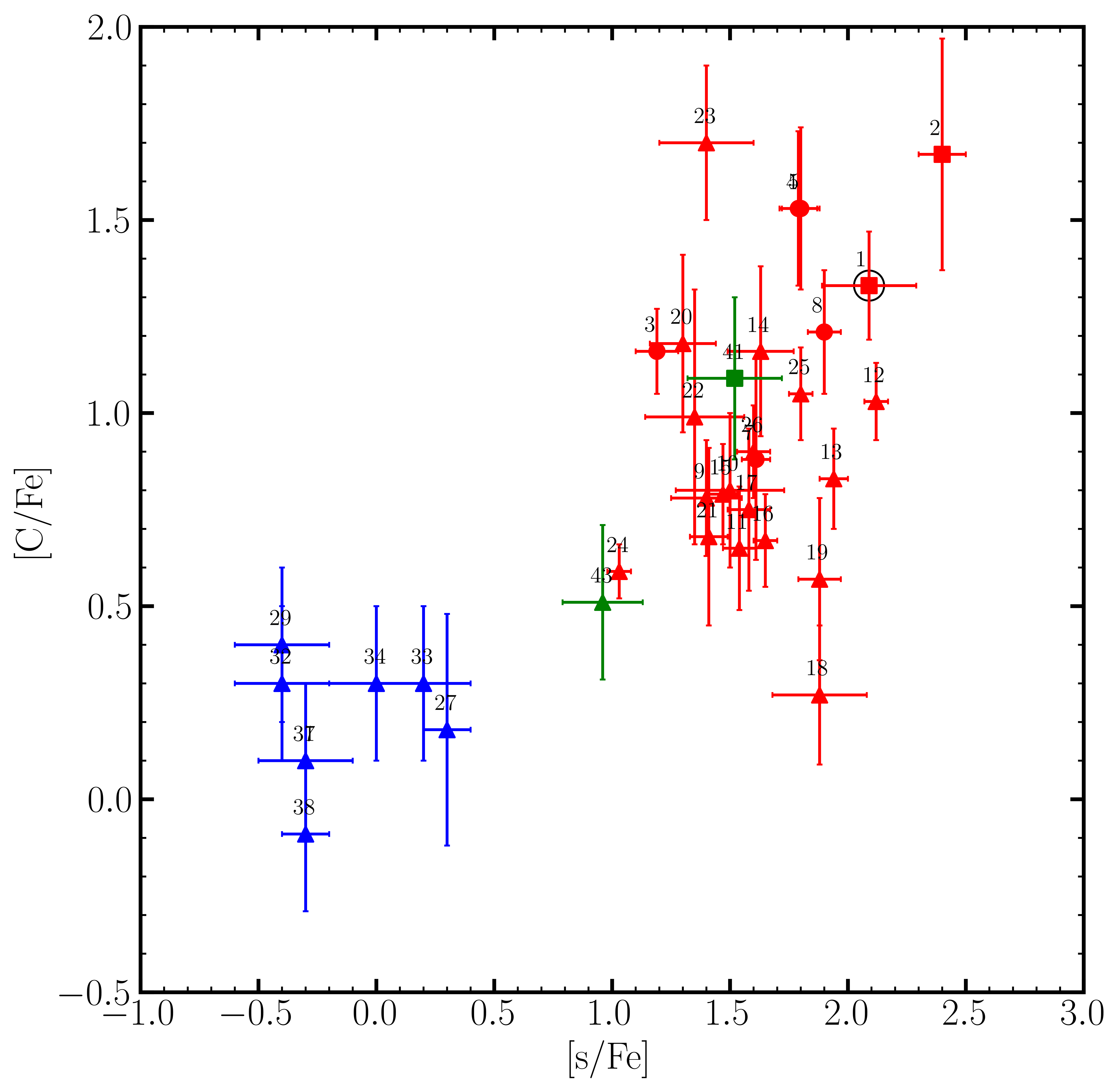}
        \caption{\cfe\ as a function of \sfe}
        \label{fig:C_sFe}
    \end{subfigure}
    \vfill
    \begin{subfigure}[b]{0.49\linewidth}
        \centering
        \includegraphics[width=\linewidth]{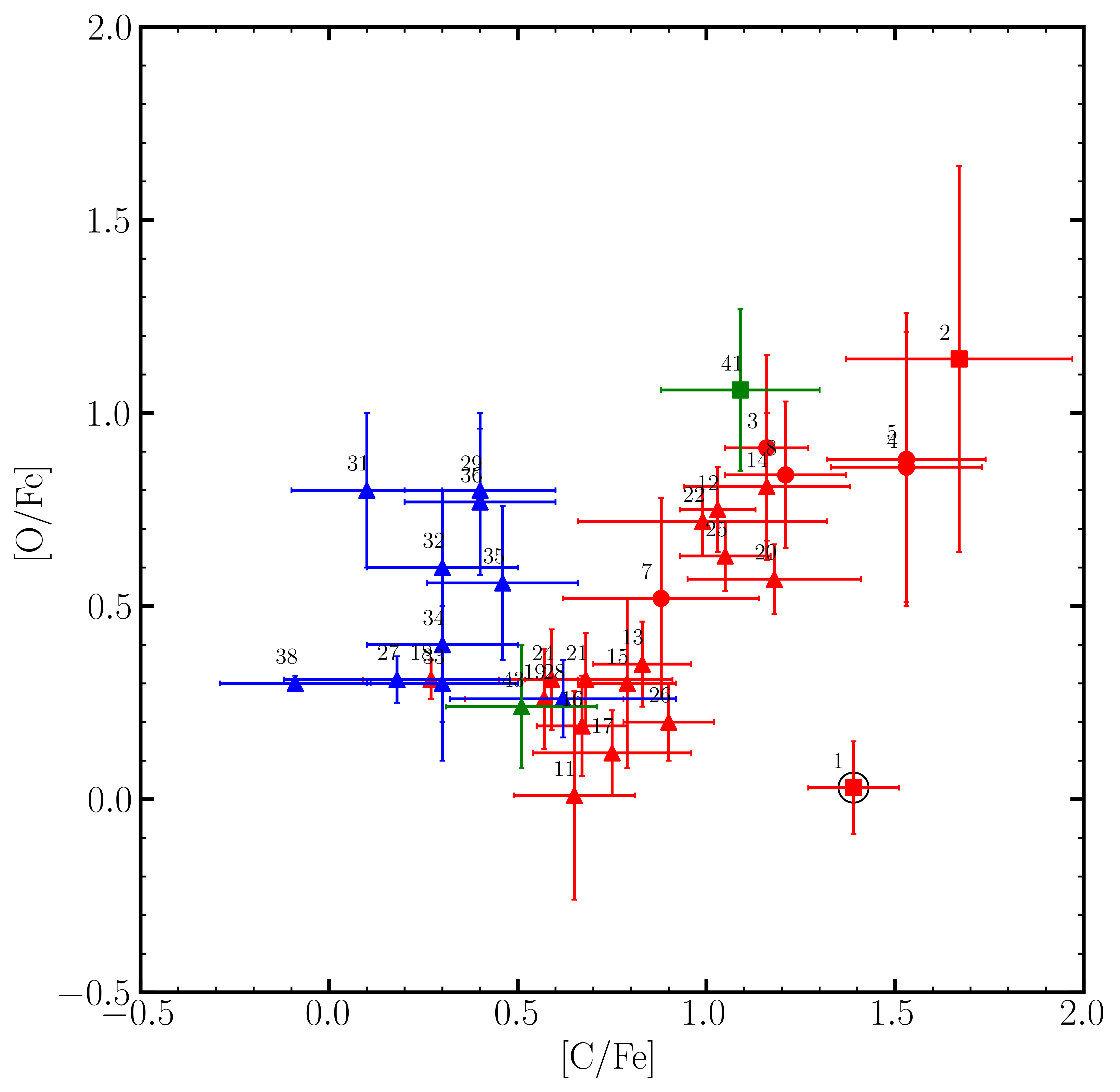}
        \caption{\cfe\ as a function of \ofe.}
         \label{fig:C_O}
    \end{subfigure}
    \hfill
    \begin{subfigure}[b]{0.49\linewidth}
        \centering
        \includegraphics[width=\linewidth]{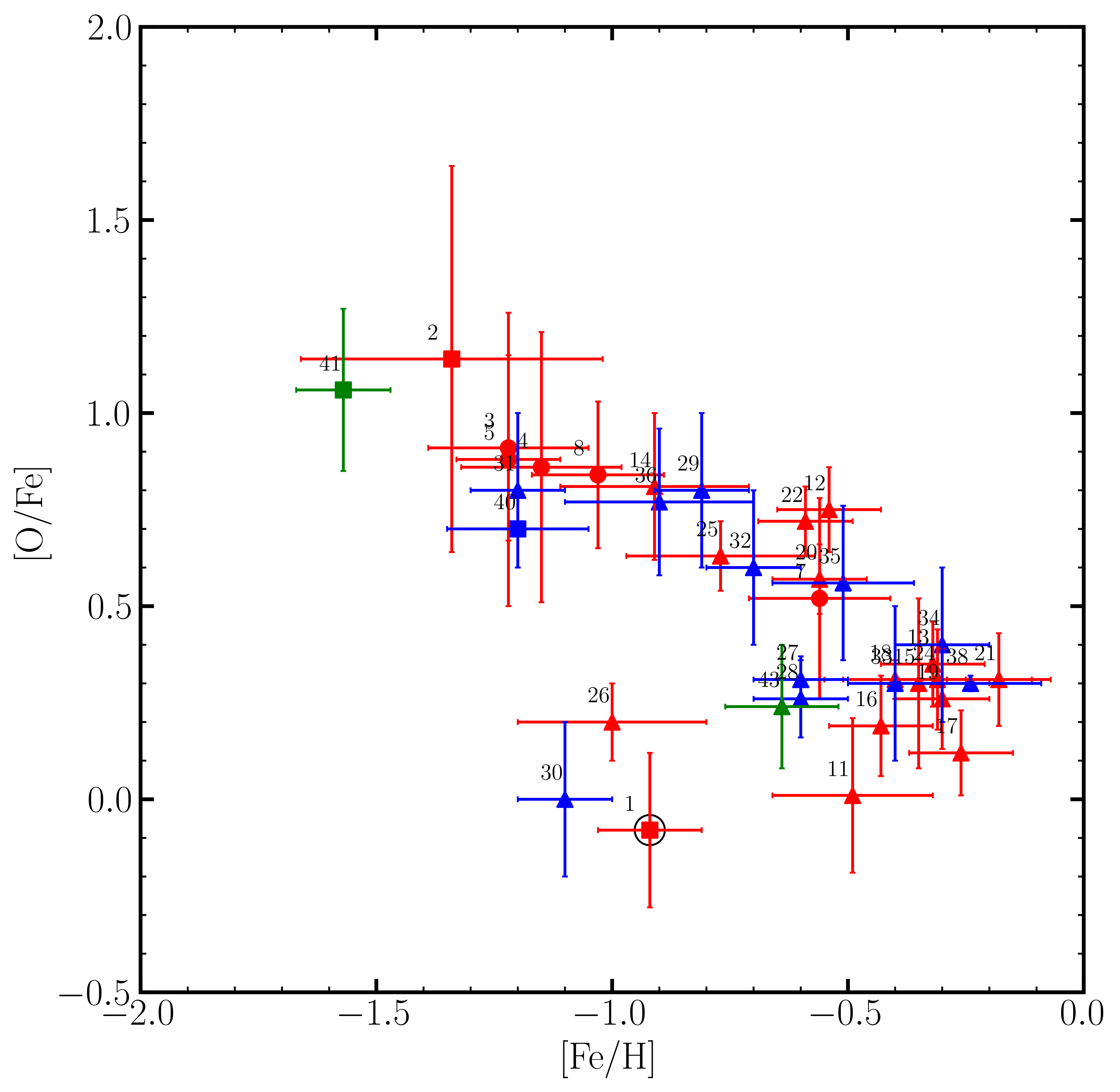}
        \caption{\ofe\ as a function of \feh.}
        \label{fig:O_Feh}
    \end{subfigure}
    \caption{Comparison of key abundance ratios for the comparative sample of post-AGB stars. Red represents single \sprocess enriched post-AGB stars, blue corresponds to single non-\sprocess enriched post-AGB stars, and green denotes binary \sprocess enriched post-AGB stars. The target J003643 is highlighted with a black circle. LMC stars are shown as circles, SMC stars as squares, and Galactic stars as triangles. The numbers correspond to the index values listed in Table~\ref{tab:full sample}.}
\end{figure*}

\begin{figure}
    \centering
      \includegraphics[width=\linewidth]{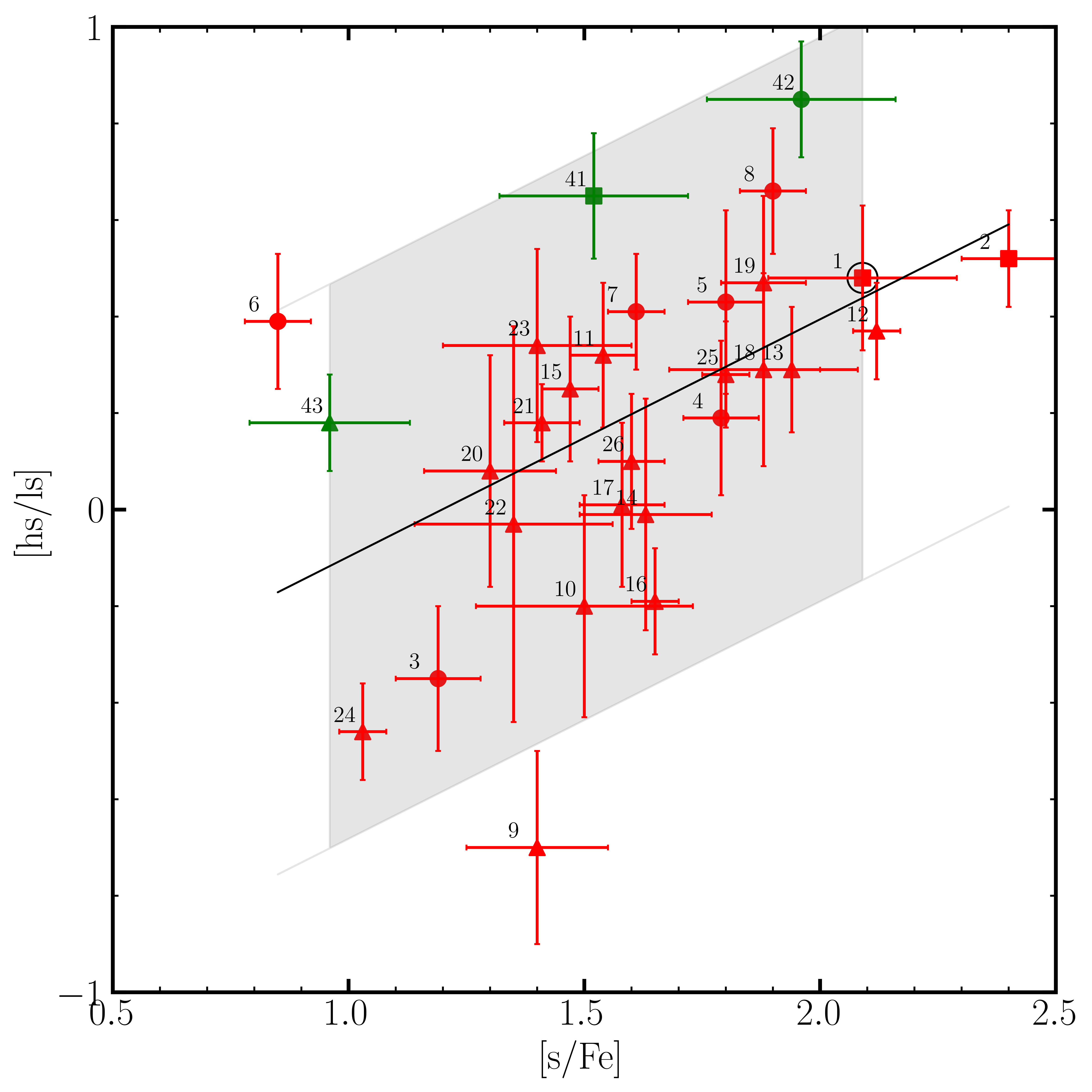}
        \caption{\hsls\ as a function of \sfe\ for the comparative sample of post-AGB stars. Colours, symbols and numbering remain the same as in Figure~\ref{fig:CO_sFe}. The black line represents the least-squares fit, while the grey-shaded region indicates the $2\sigma$ standard deviation.}
        \label{fig:sFe_hsls}
    \end{figure}
\begin{figure}
        \includegraphics[width=\linewidth]{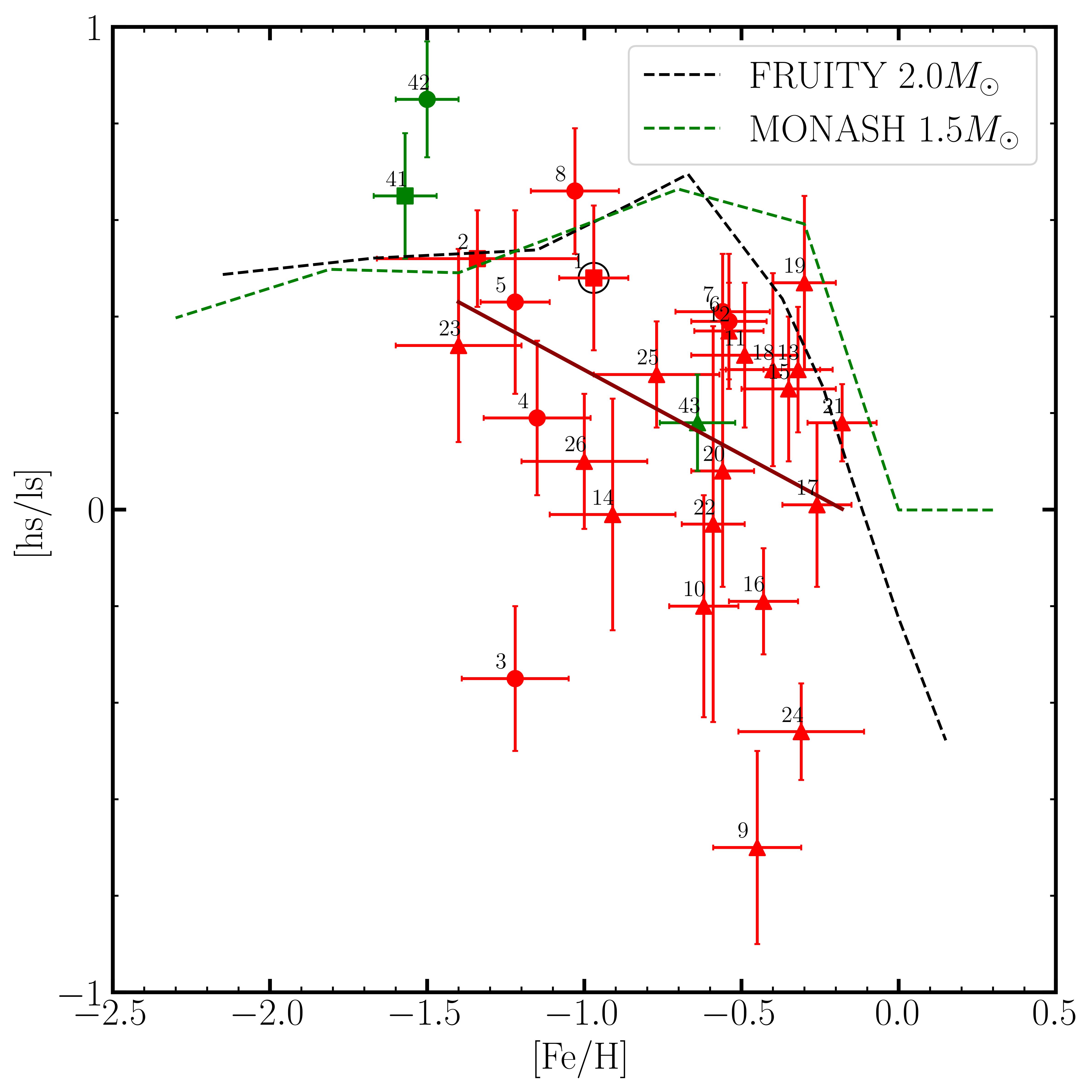}
        \caption{\hsls\ ratio as a function of \feh\ for the comparative sample of post-AGB stars. Colours, symbols and numbering remain the same as in Figure~\ref{fig:CO_sFe}. The black dotted line represents the theoretical prediction from a 2~\Msun\ FRUITY model \citep{Cristallo2015}. The dark brown solid line represents the least-squares linear fit to the combined observational data.}
        \label{fig:FeH_hsls}
\end{figure}

Figure~\ref{fig:CO_sFe} shows the spectroscopically derived C/O ratios versus \sfe\ for the comparative sample of post-AGB stars. The data points are symbol- and colour-coded to distinguish between different groups (see caption for details). J003643 stands out as an outlier, displaying the highest C/O ratio (16.21) in the comparative sample, making it the first spectroscopically confirmed post-AGB star with such a high C/O ratio.

To investigate the high C/O ratio of J003643, we compared its \cfe--\sfe, \ofe--\cfe\ and \ofe--\feh\ distributions with the comparative sample (Figure~\ref{fig:C_sFe}, Figure~\ref{fig:C_O} and Figure~\ref{fig:O_Feh} respectively). Figure~\ref{fig:C_sFe} reveals a strong positive correlation with an increase of \cfe\ with \sfe\, consistent with expectations from TDU during the AGB phase. J003643 clearly falls along this correlation, indicating its carbon enhancement scales with its \sprocess enrichment, similar to other \sprocess enriched post-AGB stars. This suggests a typical TDU history and implies that the high C/O ratio of J003643 is not due to an unusually high carbon abundance.

Figure~\ref{fig:C_O} reveals a positive correlation between \ofe\ and \cfe, where stars with higher \cfe\ generally also exhibit higher \ofe. Similarly, Figure~\ref{fig:O_Feh} shows a negative correlation between \ofe\ and \feh, indicating that \ofe\ tends to increase with decreasing metallicity. J003643 stands out as a distinct outlier in both figures, exhibiting a relatively low $\ofe\,=-0.08\pm0.20$ dex compared to stars with similar \cfe\ and \feh. Given that oxygen is not substantially altered during the RGB or AGB phases \citep[e.g.,][]{karakas14a}, this oxygen deficit is likely a reflection of the initial chemical composition of its natal gas.

To investigate the oxygen abundance of J003643 and the comparative sample, we examined the chemical evolution trends of the Galaxy and the MCs. Chemical evolution studies of Galactic field giant stars show that \ofe\ generally increases with decreasing metallicity, reaching values up to $\sim$1 dex at $\feh\,\lesssim -1.5$ dex \citep[see, e.g.,][]{Bensby2005, Reddy2006, Mucciarelli2023}. This trend is reflected in our Galactic comparative sample, where metal-poor stars show similarly enhanced oxygen abundances and consequently C/O ratios within the typical range of $\sim$1-3. In the LMC, \ofe\ is as low as 0 dex at $\feh\,\approx$ -0.5 dex and rises to $\sim$0.5 dex at $\feh\,\approx$ -1.5 dex \citep[e.g.,][]{VanderSwaelmen2013}. Our LMC comparative sample generally follows this trend at low metallicities, though some stars with $\feh\,\lesssim-1$ dex show high \ofe\ values up to 0.8 dex (see Figure~\ref{fig:O_Feh}). This may reflect observational biases due to the limited number of low-metallicity LMC field giants available for comparison. In the SMC, field giant stars at $\feh\approx -1.0$ dex exhibit \ofe\ values ranging from $-0.1$ to $+0.4$ dex with a mean of $\approx +0.2$ dex, although with a scatter \citep[see Figure.~8 of][]{Mucciarelli2023}. At $\feh\approx -1.5$ dex, the mean \ofe\ rises to $\approx +0.8$ dex. Our SMC comparative sample, including J003643, generally follows this trend. Some stars with $\feh\,\lesssim-1.5$ dex exhibit elevated \ofe\ up to 1 dex (see Figure~\ref{fig:O_Feh}), which may again reflect observational biases from the limited number of low-metallicity SMC field giants available for comparison.

The relatively low $\ofe\,=-0.08\pm0.2$ dex, compared to the comparative sample of post‑AGB stars at similar \feh\ and \cfe, together with $[\alpha$/Fe]$\,\approx0$ dex of J003643 indicates that it likely formed from an interstellar medium (ISM) that was intrinsically oxygen-poor. Additionally, recent APOGEE observations of $\approx$3600 RGB stars in the LMC and SMC by \citet{Nidever2020} further confirm both $[\alpha$/Fe] and \ofe\ remain essentially solar at $\feh\, \approx$ -1, demonstrating that oxygen‑poor, $\alpha$‑solar abundance patterns are indeed present in SMC star‑forming regions. Therefore, the relatively high C/O ratio observed in J003643 primarily reflects formation from oxygen-poor gas, rather than resulting from an unusually high carbon abundance.

Having established the likely chemical origin of J003643’s high C/O ratio, we next examined its \sprocess efficiency in the context of the comparative sample, using the \hsls--\sfe\ plane. Figure~\ref{fig:sFe_hsls} shows the spectroscopically derived \hsls\ ratios plotted against \sfe\ for the comparative sample. Previous studies \citep{Vanwinckel2000, Reyniers2004, Vanaarle2013} have reported a positive correlation between \sfe\ and \hsls, where stars with higher \sfe\ tend to exhibit higher \hsls\ ratios, reflecting a shift in the \sprocess path toward heavier elements at higher neutron exposures. Our analysis confirms this correlation and supports the theoretical expectation that stronger \sprocess enrichment corresponds to a greater relative production of hs elements compared to ls elements, due to increased neutron flux per iron seed. However, there is a $2\sigma$ scatter. This is unlikely to arise from observational uncertainties, which are relatively well constrained, but may reflect variations in neutron exposure, the structure and extent of the \textsuperscript{13}C pocket, initial stellar mass, or other processes not fully captured by current models. Additionally, J003643 fits well within the least-squares relation derived from the comparative sample. This indicates that its heavy-to-light \sprocess element production is typical for its overall \sfe\ value. Furthermore, as in the C/O–\sfe\ plane, no systematic trend is observed with galactic environment, as stars from the Galaxy, LMC, and SMC remain evenly distributed.

Lastly, to examine how neutron exposure varies with stellar metallicity, Figure~\ref{fig:FeH_hsls} presents the spectroscopically derived \hsls\ ratio as a function of \feh\ for the comparative sample. The colours and symbols follow the same scheme as in Figure~\ref{fig:CO_sFe}, and the black and green dotted line shows the theoretical prediction from a 2~~\Msun\ FRUITY model \citep{Cristallo2015} and 1.5~~\Msun\ MONASH model \citep[see]{Karakas2014, karakas16, Karakas2018} respectively. Previous studies \citep[e.g.,][]{desmedt2016} did not report a clear correlation between \feh\ and \hsls, likely due to the limited number ($\sim$14) of well-characterised post-AGB stars. However, with the inclusion of a more comprehensive sample of \sprocess enriched stars ($\sim$29), we now find a negative correlation between \feh\ and \hsls. To explore the significance of this correlation, we performed a Pearson correlation test. Although the test gives a statistically significant negative correlation between \feh\ and \hsls\ (slope=$-0.35$, $r=-0.42$, $p=0.024$), the object‑to‑object scatter is large, with objects 3,4, and 29 falling outside the $2\sigma$ errors. Additionally, a significant fraction of the current sample does not match the model predictions. Although this trend supports theoretical predictions that lower-metallicity stars have higher neutron-to-seed ratios, favouring the production of heavier \sprocess elements and yielding elevated \hsls\ ratios \citep[see][]{gallino98, Busso2001}, a more complete sample is required to confirm the correlation. Additionally, J003643 agrees with the model prediction, further reinforcing that its \sprocess efficiency is consistent with expectations for its metallicity.  However, the persistent disagreement between model predictions and observations for the majority of stars, especially at \feh$\,\lesssim-0.5$ dex, suggests that factors beyond metallicity, such as initial mass, TDU efficiency, dominant neutron source (\textsuperscript{13}C vs. \textsuperscript{22}Ne), and mass-loss history, may significantly influence the resulting \hsls\ ratios. This disagreement does not appear to depend on galactic environment, as stars from the Galaxy, LMC, and SMC are distributed uniformly in the \hsls--\feh\ plane.

\subsection{Understanding the Nucleosynthetic origin of J003643 Using Predictions from Different Stellar Evolutionary Models}
\label{sec:model_disc}
To investigate the nucleosynthetic origin of the heavy-element enrichment observed in J003643, we first assess whether its chemical composition could reflect pre-enrichment from its birth environment in the SMC. The SMC’s chemical evolution is distinct from that of the Galaxy, characterised by low metallicities and modest enhancements in \sprocess elements. For example, \citet{Mucciarelli2023} report that the average [Zr/Fe], [Ba/Fe], and [La/Fe] abundances in SMC field giants remain below $\sim$0.5 dex, even at low metallicities. In contrast, J003643 exhibits significantly higher values for these \sprocess indicators, pointing toward substantial intrinsic enrichment during its AGB phase. Additionally, studies of SMC field stars and clusters have revealed signs of enhanced \rprocess contributions, particularly in elements like europium (Eu), with [Eu/Fe] often exceeding typical Galactic values at comparable metallicities \citep[e.g.,][]{Reggiani2021, Mucciarelli2023B}.

\begin{figure}
    \centering
    \includegraphics[width=\linewidth]{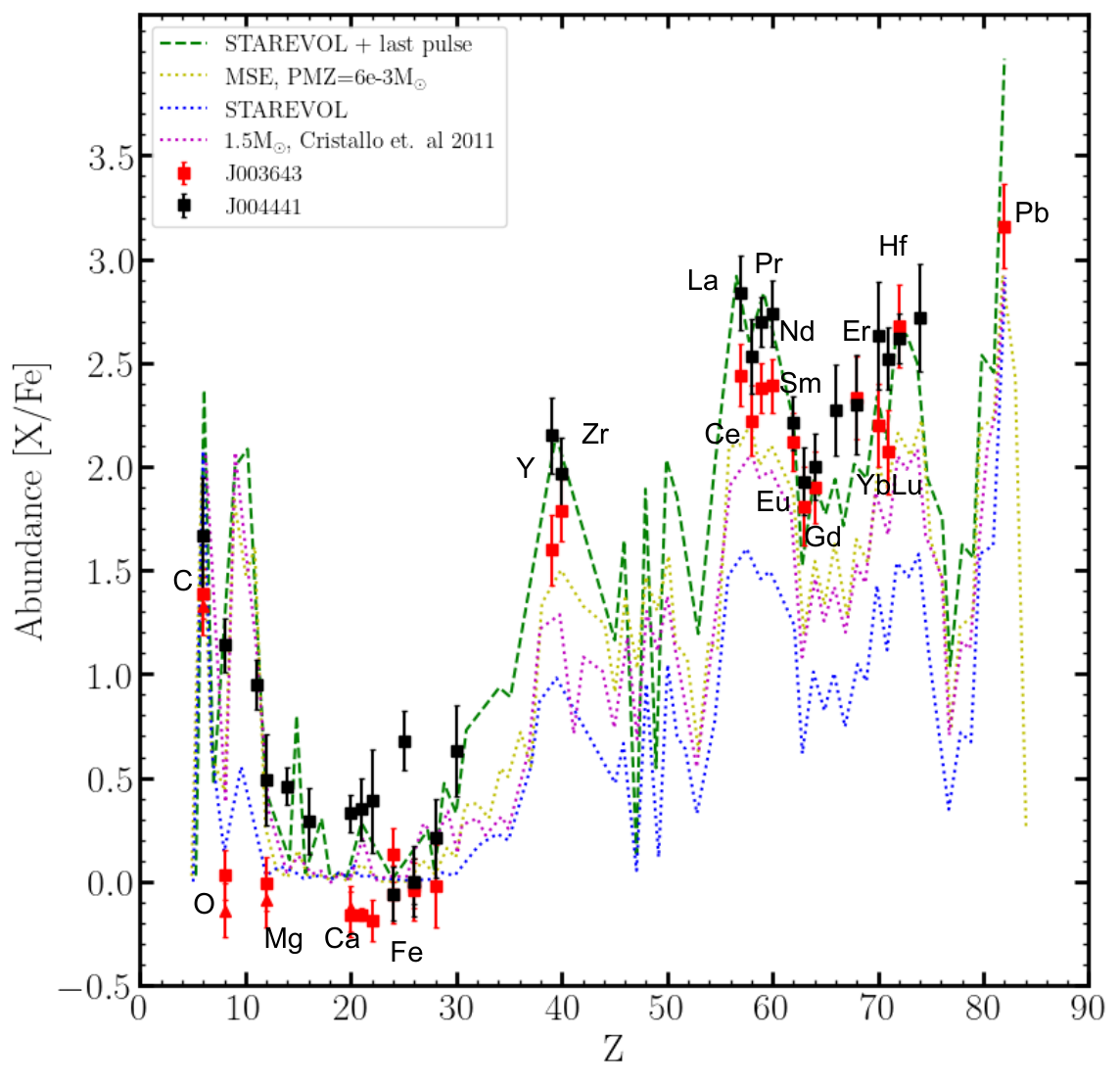}
    \caption{Comparison of the spectroscopically derived abundances of J003643 with stellar evolutionary model predictions reproduced from \citet{desmedt12}, originally used for the post-AGB star J004441. Selected elements are labelled for reference. See text for details.}
    \label{fig:comp_lit}
\end{figure}

To further explore the intrinsic nucleosynthesis, we compared the spectroscopically derived abundances of J003643 with predictions from various stellar evolutionary models previously applied to the most \sprocess enriched post-AGB star, J004441 \citep[see][ references therein]{desmedt12}. As shown in Figure~\ref{fig:comp_lit}, adapted from \citet{desmedt12}, the best-fitting model for J004441 is the STAREVOL model with a late thermal pulse. However, despite J003643 having similar stellar parameters and \sfe, this model fails to reproduce its observed abundances, particularly for C, O, several hs and ls elements, and most notably Pb. Other models considered in the same study similarly fail to simultaneously match the full abundance pattern of J003643.

\begin{figure}
    \centering
    \includegraphics[width=\columnwidth]{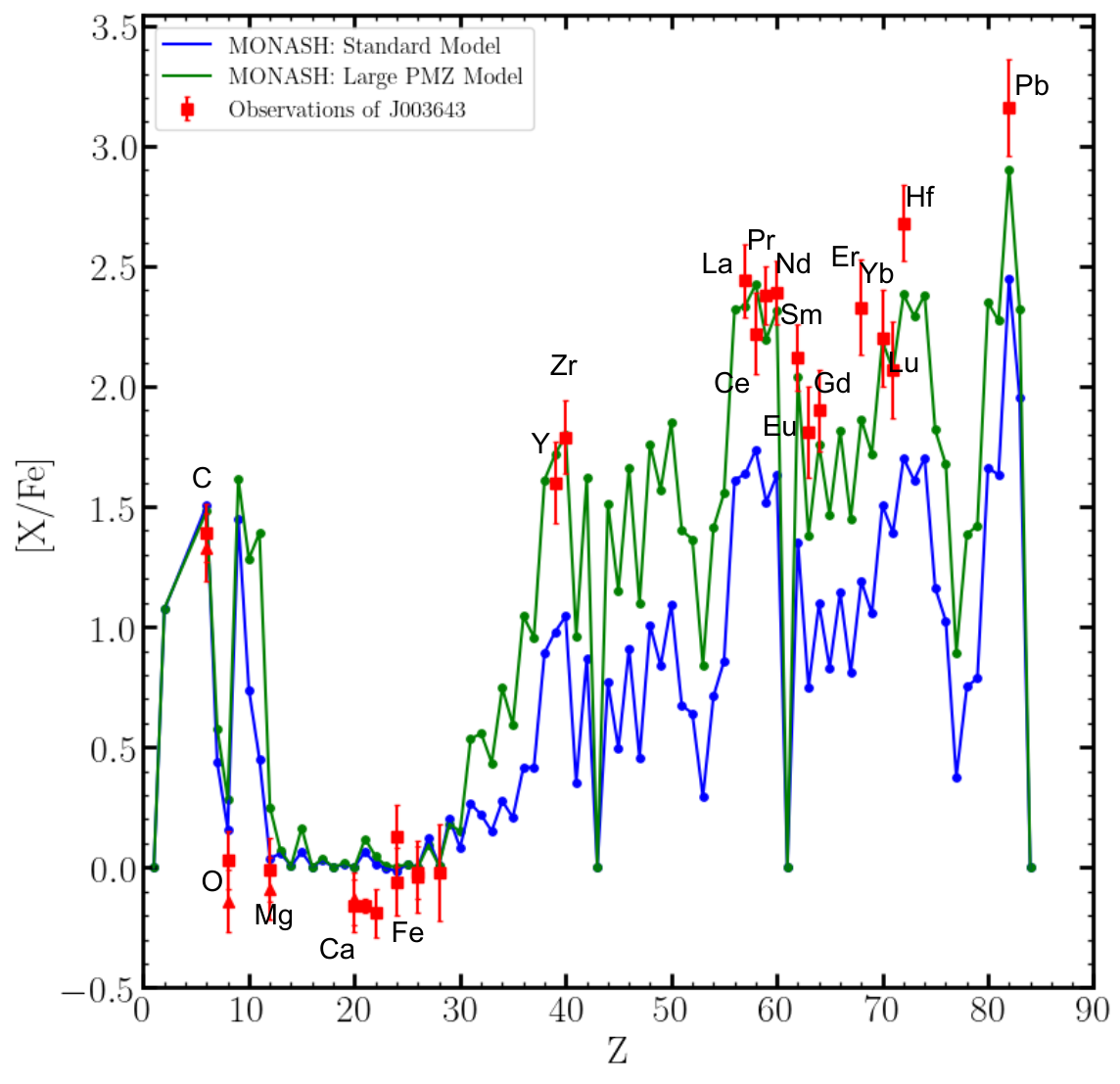}
    \caption{Comparison of the spectroscopically derived abundances of J003643 with predictions from the MONASH stellar evolution code. The y-axis represents the elemental abundance ratio \xfe, and the x-axis denotes the atomic number Z. The red squares and triangles with error bars show the LTE and NLTE-corrected abundances of J003643 derived in this study (see Section~\ref{sec:abund_analysis}). The blue line with dots corresponds to the predicted abundance pattern from the standard MONASH model with no convective overshoot, with an initial mass of 1.7~\Msun\ and metallicity Z = 0.0014. The green line with dots corresponds to the predicted abundance pattern from the MONASH model with an initial mass of 1.7~\Msun\ and metallicity Z = 0.0014, incorporating a large PMZ of 0.006~\Msun. Some elements are labelled for reference. See text for details.}
    \label{fig:MONASH}
\end{figure}
To perform a more targeted analysis of the heavy-element abundance of J003643, we strategically employed two sets of stellar evolution and stellar evolutionary models: the MONASH (Section~\ref{sec:monash_models}) and FRUITY (Section~\ref{sec:fruity_models}) grids. 
\begin{figure}
    \centering
    \includegraphics[width=\linewidth]{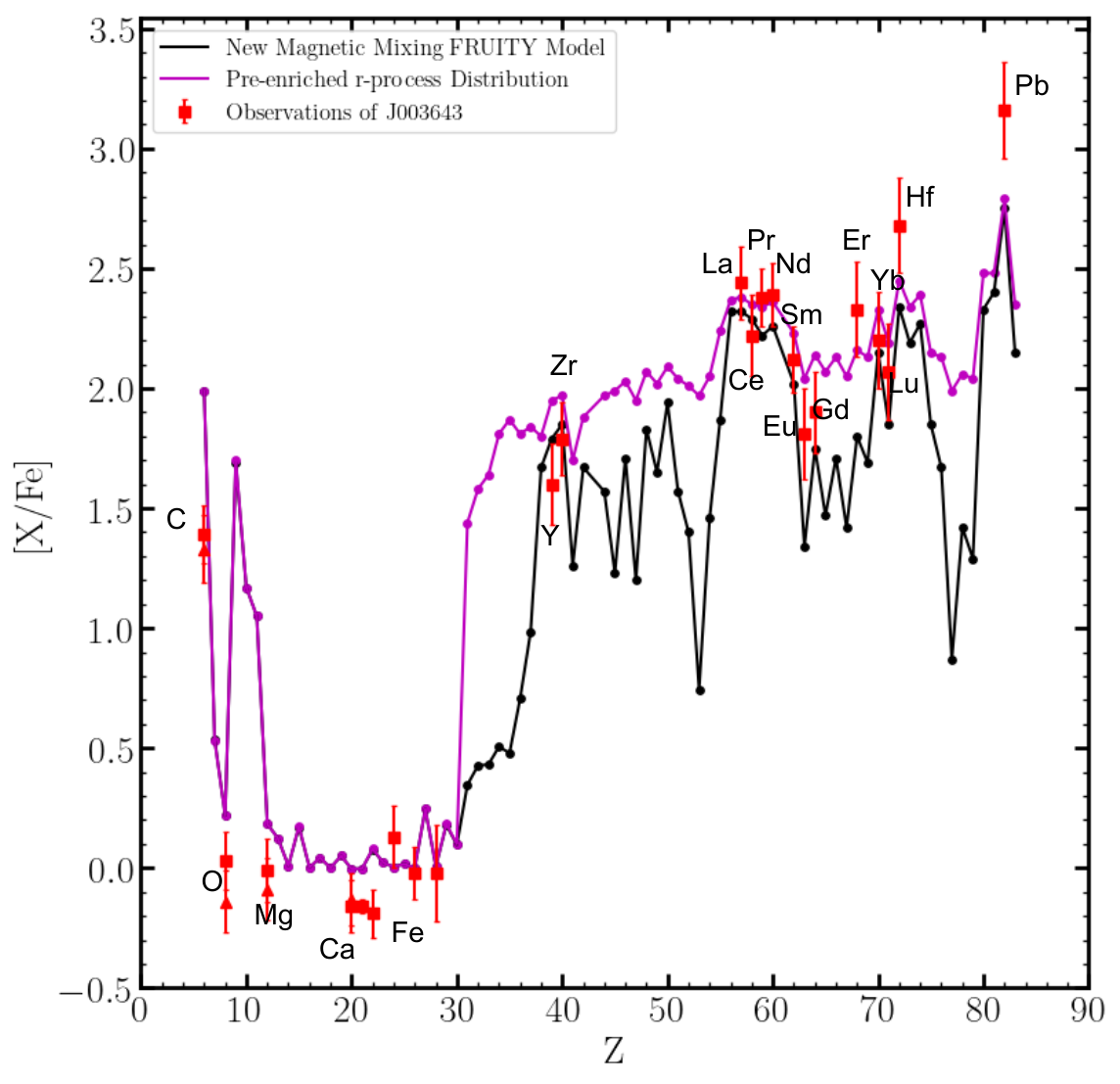}
    \caption{Comparison of the spectroscopically derived abundances of J003643 with theoretical nucleosynthesis predictions from the new magnetic mixing FRUITY model. The y-axis represents the elemental abundance ratio \xfe, and the x-axis denotes the atomic number Z. The red squares and triangles with error bars show the LTE and NLTE-corrected abundances of J003643 derived in this study (see Section~\ref{sec:abund_analysis}). The black line corresponds to the predicted abundance pattern from the new magnetic mixing FRUITY model with an initial mass of 2~\Msun\ and metallicity Z = 0.001. The magenta line represents a version assuming a pre-enriched \rprocess\ component with [r/Fe] = 2 dex, applied from Ga to Bi. Some elements are labelled for reference. See text for details.}
    \label{fig:fruity}
\end{figure}

\begin{figure}
    \centering
    \includegraphics[width=\linewidth]{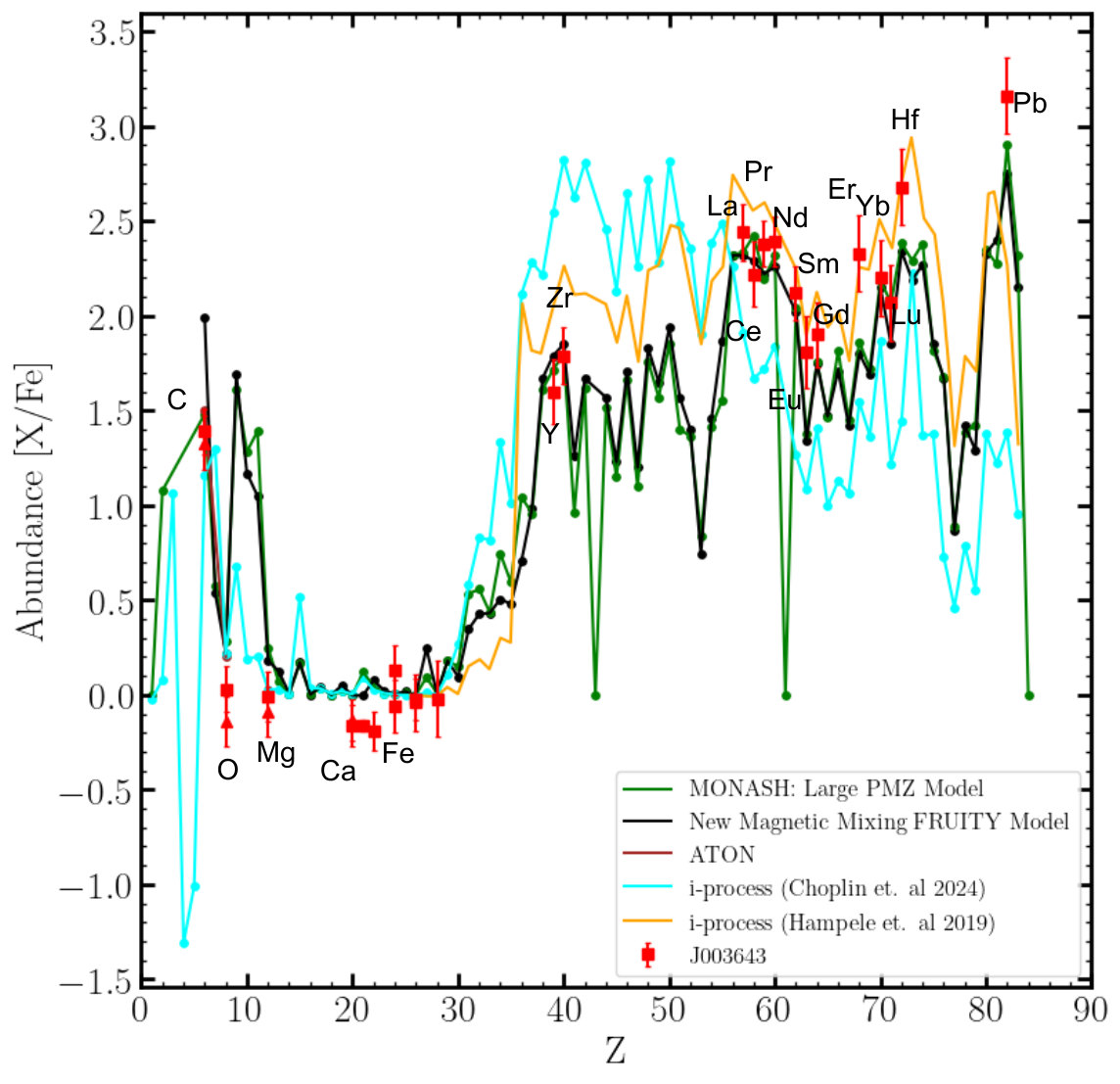}
    \caption{Comparison of the spectroscopically derived abundances of J003643 with theoretical nucleosynthesis predictions from various stellar evolutionary models. The y-axis represents the elemental abundance ratio \xfe, and the x-axis denotes the atomic number Z. The red squares and triangles with error bars show the LTE and NLTE-corrected abundances of J003643 derived in this study (see Section~\ref{sec:abund_analysis}). The green and black lines correspond to predictions from the best-fitting MONASH and FRUITY models, respectively, while the magenta line represents a version assuming a pre-enriched \rprocess\ component with [r/Fe] = 2 dex, applied from Ga to Bi. Additional comparisons include the ATON model (brown line; \citep{ventura98}), an \sprocess model (yellow line; \citep{Lugaro2015}), and \iprocess models (cyan \citet{Choplin2024} and orange lines; \citet{Hampel2019}). Some elements are labelled for reference. See text for details.}
    \label{fig:models}
\end{figure}

\textbf{MONASH:} The standard MONASH model (see Section~\ref{sec:monash_models} and Figure~\ref{fig:MONASH}; blue line) significantly underpredicts the derived heavy-element abundances of J003643. However, the MONASH model, assuming a large PMZ of 0.006~\Msun\ (see Section~\ref{sec:monash_models} and Figure~\ref{fig:MONASH}; green line), predicts an \sprocess distribution that is very similar to the derived abundance of J003643, particularly for Fe-peak, the light \sprocess (ls) peak (Y, Zr) and the second peak (Ce, Nd). Despite this improvement, the predicted Pb abundance remains lower than derived. Additionally, elements such as Er and Hf are also underpredicted. The model attains a final C/O\,= 12.5, slightly overestimating the oxygen abundance.

\textbf{FRUITY:}The new magnetic mixing FRUITY model (see Section~\ref{sec:fruity_models} and Figure~\ref{fig:fruity}; black line) slightly overestimates the abundances of CNO while the Fe-peak, ls and second \sprocess peak elements are well reproduced. The FRUITY model also underpredicts the derived Pb abundance. Additionally, it underpredicts the derived abundances of Eu, Er, Lu, Yb, and Hf. To explore this, we considered a modified version of the model that includes an initially \rprocess enhanced composition, adopting a solar \rprocess distribution as prescribed by \citet{Prantzos2020}, scaled by +2 dex ([r/Fe] = +2 dex) from Ga to Bi (see Figure~\ref{fig:fruity}, magenta line). While the pre-enrichment scenario improves the fit for several \rprocess sensitive elements, particularly Eu, Er, and Hf, it does not broadly affect the overall abundance pattern. This raises the possibility that J003643 may have formed from a locally \rprocess enriched ISM. This is supported by studies of the SMC, where field stars at low metallicities show systematically elevated [Eu/Fe] ratios compared to their Galactic counterparts, indicating a more efficient early \rprocess enrichment in the SMC \citep{Reggiani2021, Mucciarelli2023}. Although the average [Eu/Fe] in the SMC remains below +0.5 dex, the possibility of localised enrichment episodes, e.g., from neutron star mergers, remains plausible.

\textbf{\iprocess Models: Additionally, studies by \citet{Hampel2019} suggest that the abundance patterns of some Pb‑poor post‑AGB stars (\pbfe\,$<2.5$ dex) may be better explained by \iprocess\ nucleosynthesis \citep{Cowan1977, Malaney1986, Hampel2016}. At intermediate neutron densities, the \iprocess boosts the ls and hs elements without producing large amounts of Pb. Although J003643 is not Pb‑poor, we tested whether \iprocess nucleosynthesis could nonetheless reproduce its observed abundance pattern.} The \iprocess model of \citet{Choplin2024}, computed for a 1.5~\Msun, Z = 0.001 AGB star (Figure~\ref{fig:models}; cyan line), substantially overestimates the abundances of the ls elements while severely underpredicting the hs elements, including Pb. Similarly, the best-fitting \iprocess model from \citet{Hampel2019}, assuming a neutron density of $n\,= 10^{11}$ cm$^{-3}$ (Figure~\ref{fig:models}; orange line), overestimates many of the heavy elements but still fails to reproduce the observed Pb abundance. These discrepancies suggest that \iprocess nucleosynthesis is unlikely to account for the observed abundance pattern of J003643.

Figure~\ref{fig:models} summarises the best-fitting models, including ATON, MONASH, and FRUITY, as well as the \iprocess\ models from \citet{Hampel2019} and \citet{Choplin2024} for comparison. Overall, the comparison between the observed abundances of J003643 and predictions from various stellar evolutionary models indicates that its heavy element enrichment results from an intrinsic \sprocess nucleosynthesis involving efficient TDU. However, the observed Pb abundance remains higher than predicted by all current models, despite varying nucleosynthetic assumptions. This persistent discrepancy highlights a gap in our understanding of Pb production in AGB stars. As the first post-AGB star with a direct and precise Pb abundance measurement, J003643 offers a valuable benchmark for refining theoretical models of heavy-element nucleosynthesis.

\subsection{Investigating the Pb Abundances and the Longstanding Pb Discrepancy in Post-AGB Stars}
\label{sec:lead}
\input{Table7}

Pb is expected to be a major product of the \sprocess in low-mass AGB stars ($<3~\Msun$), particularly at low metallicities where the neutron exposure is higher \citep[e.g.,][]{gallino98, Goriely2000}. As the endpoint of the \sprocess chain, theoretical models predict that Pb should be strongly overabundant relative to the ls and hs peaks, resulting in high \pbhs\ and \pbls\ ratios \citep[e.g.,][]{Busso2001, Lugaro2012}. Accurate Pb measurements are thus essential to constrain the efficiency and termination of the \sprocess in these stars.

Despite these predictions, observational studies of \sprocess enriched post-AGB stars have consistently failed to detect Pb at the expected levels. In most cases, only upper limits have been obtained using SSF, due to the weakness of Pb lines in the optical, blending issues, and uncertainties in atomic data \citep[e.g.,][]{desmedt2016, Kamathuniverse2021}. Depending on stellar temperature, Pb is probed using either the Pb I 4057.807~\AA\ line or the Pb II 5608.853~\AA\ line. Upper limits of derived Pb abundance for both single and binary \sprocess enriched post-AGB stars, compiled from the literature, are summarised in Table~\ref{tab:lead}. For J003643, however, we derive the first direct Pb abundance measurement.

Previous studies have shown that for post-AGB stars with low metallicities (\feh\,$<-0.5$ dex), the derived \pbhs\ and \pbls\ upper limits were significantly lower than theoretical predictions, with the discrepancy growing at lower metallicities \citep[see][]{Vanwinckel2000, desmedt12, deSmedt2015, desmedt2016}. This persistent disagreement between predicted and deduced Pb abundances remains one of the key unresolved challenges in \sprocess nucleosynthesis \citep{Kamathuniverse2021}.

Proposed explanations for this discrepancy include incomplete or inaccurate atomic data for Pb lines, limitations in spectral modelling, poor S/N in the blue region near the Pb I line, and incomplete treatment of physical processes governing the \sprocess in low-mass stars \citep[e.g.,][]{Lugaro2012, karakas16, desmedt2016}. Additional effects—such as rotation, internal mixing (e.g., gravity-wave induced), and enhanced mass loss during or prior to the thermally-pulsing AGB phase—may add further uncertainty to the results \citep[e.g.,][]{Herwig2005, Siess2004, karakas14a, Piersanti2013}.

NLTE effects may also contribute to the discrepancy, particularly in warm, low-gravity post-AGB stars. As discussed in \citet{Kamathuniverse2021}, NLTE calculations by \citet{2012A&A...540A..98M} for the Pb I line at 4057.807~\AA\ in the Sun and a sample of metal-poor stars ($-2.95\,<\,\feh\,<\,-0.7$) showed that NLTE effects lead to systematically lower line absorption, resulting in positive abundance corrections of $\Delta_{\mathrm{NLTE}}\sim$0.2–0.45 dex. These effects become more pronounced at lower metallicities. However, the calculations were performed for stars with lower $T_{\mathrm{eff}}$ and higher surface gravity than typical post-AGB stars. 

In this study, we perform an exploratory NLTE calculation using a simplified toy model for neutral Pb I to estimate NLTE effects on Pb abundance determinations. We focus on the neutral Pb I line, as Pb II—the dominant ionisation stage—is expected to remain close to LTE due to its stronger coupling with the local radiation field, making NLTE effects less significant \citep{2012A&A...540A..98M}.

As in Section~\ref{sec:nlte}, the NLTE calculations were performed using \texttt{Balder} \citep{amarsi2018Balder}, based on \texttt{Multi3D} \citep{Multi3D2009}. The calculations were performed on a grid of MARCS model atmospheres (see \citealt{amarsi2020NLTEgalah} for details), for a range of microturbulence values ($1$, $2$, and $5\,\mathrm{km\,s^{-1}}$) and \pbfe\ (from $-1.5$ to $+3.5$ in steps of $0.5$ dex). The abundance corrections were interpolated or extrapolated (taking the edge values) onto the stellar parameters of interest. 

For this test, a simple ``toy'' model atom was constructed for neutral lead (Pb I). Energy levels and bound-bound transition rates were taken from the National Institute of Standards and Technology Atomic Spectra Database (NIST ASD; \citealt{2020Atoms...8...56R}). Photoionisation cross-sections were assumed to be hydrogenic. Inelastic electron collisions were estimated using the recipe of \citet{1962ApJ...136..906V} for excitation and \citet{1973asqu.book.....A} for ionisation. Inelastic hydrogen collisions were estimated using the Drawin recipe \citet{1993PhST...47..186L}; hydrogen collisional ionisation was assumed to be negligible \citep{2011A&A...530A..94B}. For the collisional excitation recipes, $\log gf$ was assumed to be $-1.5$ for permitted lines (if missing from NIST ASD), and $-3$ for forbidden lines.

This toy model is somewhat simpler than that used in \citet{2012A&A...540A..98M}. Possibly the main limitation here relative to that work is the absence of bound-bound transitions of high excitation potential (where \citealt{2012A&A...540A..98M} include these via their own calculations using the Cowan code). Nevertheless, the resulting abundance corrections for the non-detected and hence upper limit of the Pb I 4057.807~\AA\ line are of the same order of magnitude as those presented in their Table 1, taking into account different assumptions about the inelastic hydrogen collisions. This could be because the NLTE effects are controlled by the photoionisation transitions as well as the radiative pumping of UV lines, which are included in our toy model. The \pbfe\ after applying the NLTE corrections for the post-AGB stars are presented in the last column of Table~\ref{tab:lead}.

\begin{figure*}
    \centering
    \begin{subfigure}[b]{0.49\linewidth}
        \centering
        \includegraphics[width=\linewidth]{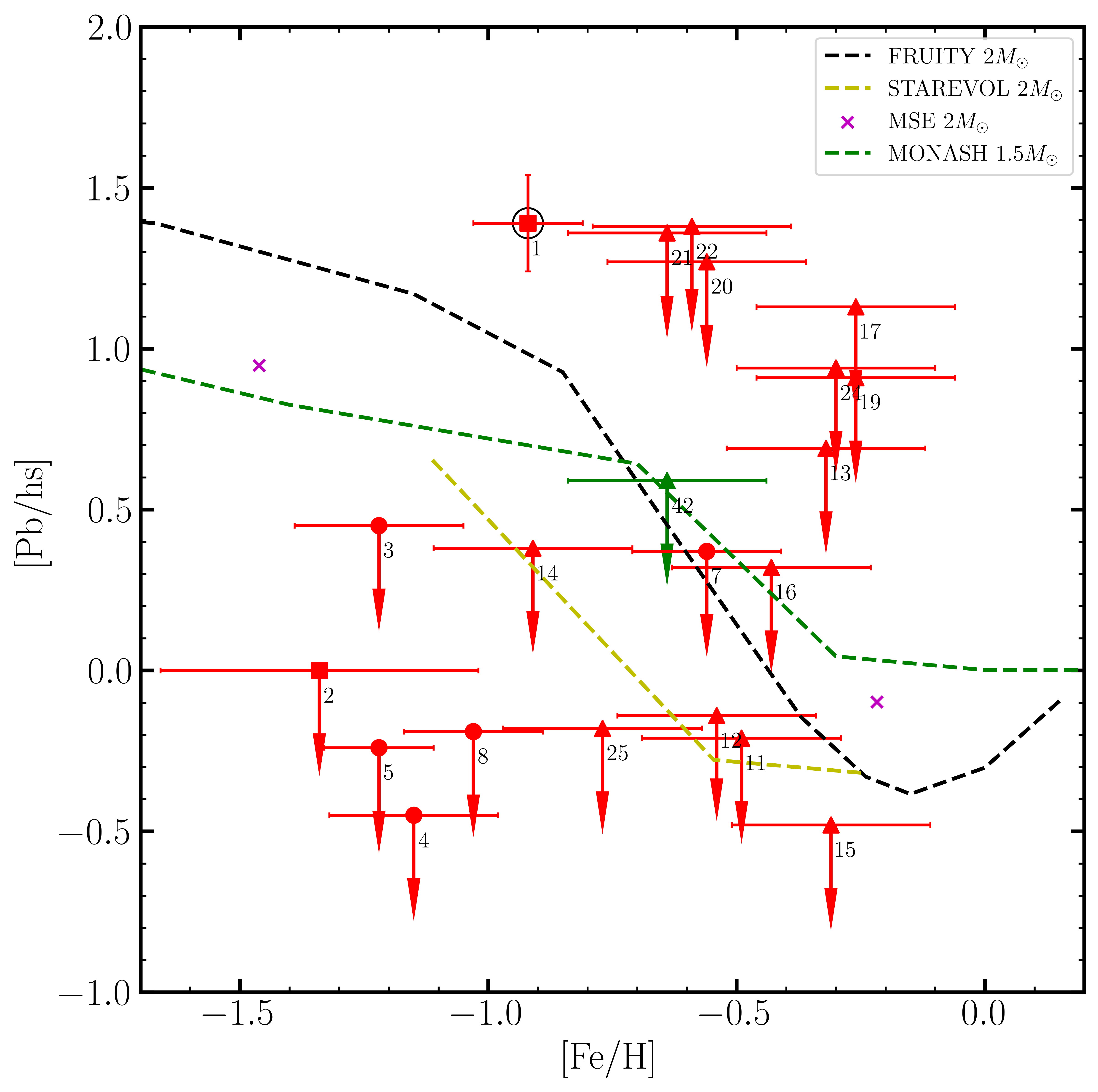}
    \end{subfigure}
    \hfill
    \begin{subfigure}[b]{0.49\linewidth}
        \centering
        \includegraphics[width=\linewidth]{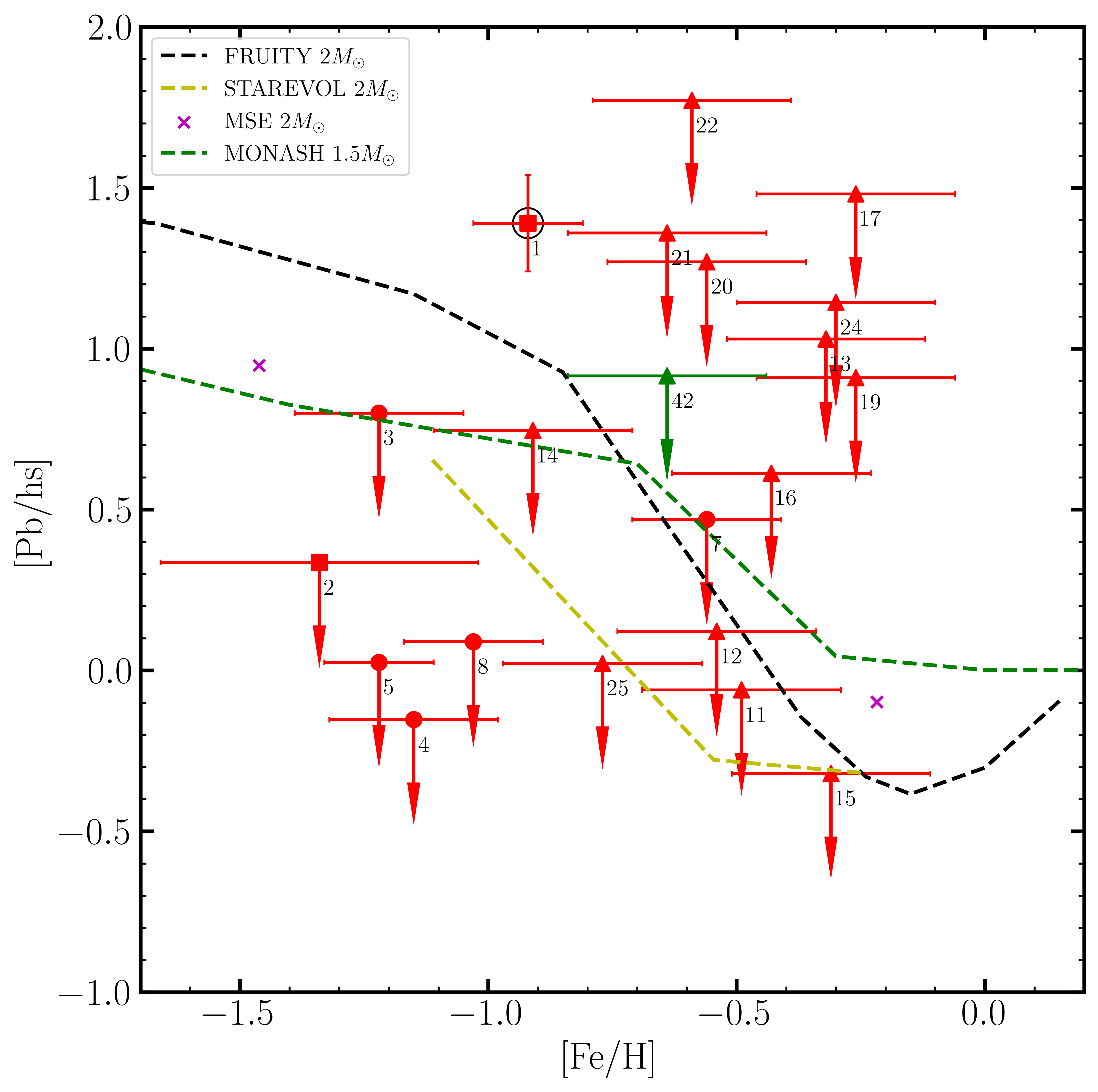}
    \end{subfigure}
    \caption{Comparison of \pbhs\ versus \feh\ for post-AGB stars with and without NLTE corrections. The left panel shows LTE-based upper limits of Pb abundances, while the right panel includes NLTE-corrected upper limits of Pb abundances derived in this study. Red symbols represent single \sprocess enriched post-AGB stars, while green symbols indicate binary \sprocess enriched post-AGB stars. LMC stars are shown as circles, SMC stars as squares, and Galactic stars as triangles. Downward arrows denote upper limits for all stars except J003643. The error bars along the x-axis indicate uncertainties in \feh. Theoretical predictions from FRUITY, STAREVOL, MSE, and MONASH stellar evolutionary models are overlaid for reference. See text for more details.}
    \label{fig:hsNLTE_lead}
\end{figure*}

\begin{figure*}
    \begin{subfigure}[b]{0.49\linewidth}
        \centering
        \includegraphics[width=\linewidth]{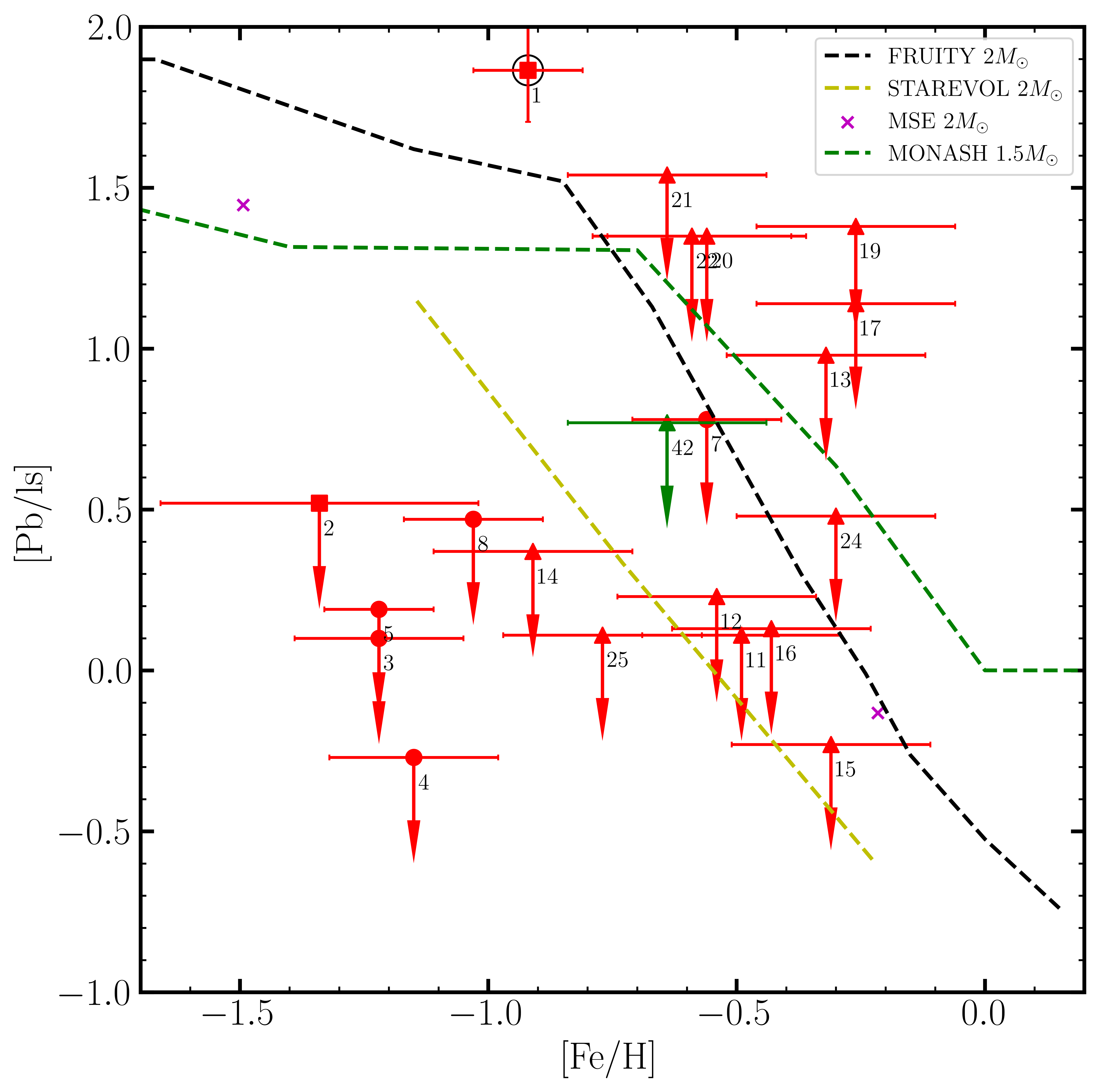}
    \end{subfigure}
    \hfill
    \begin{subfigure}[b]{0.49\linewidth}
        \centering
        \includegraphics[width=\linewidth]{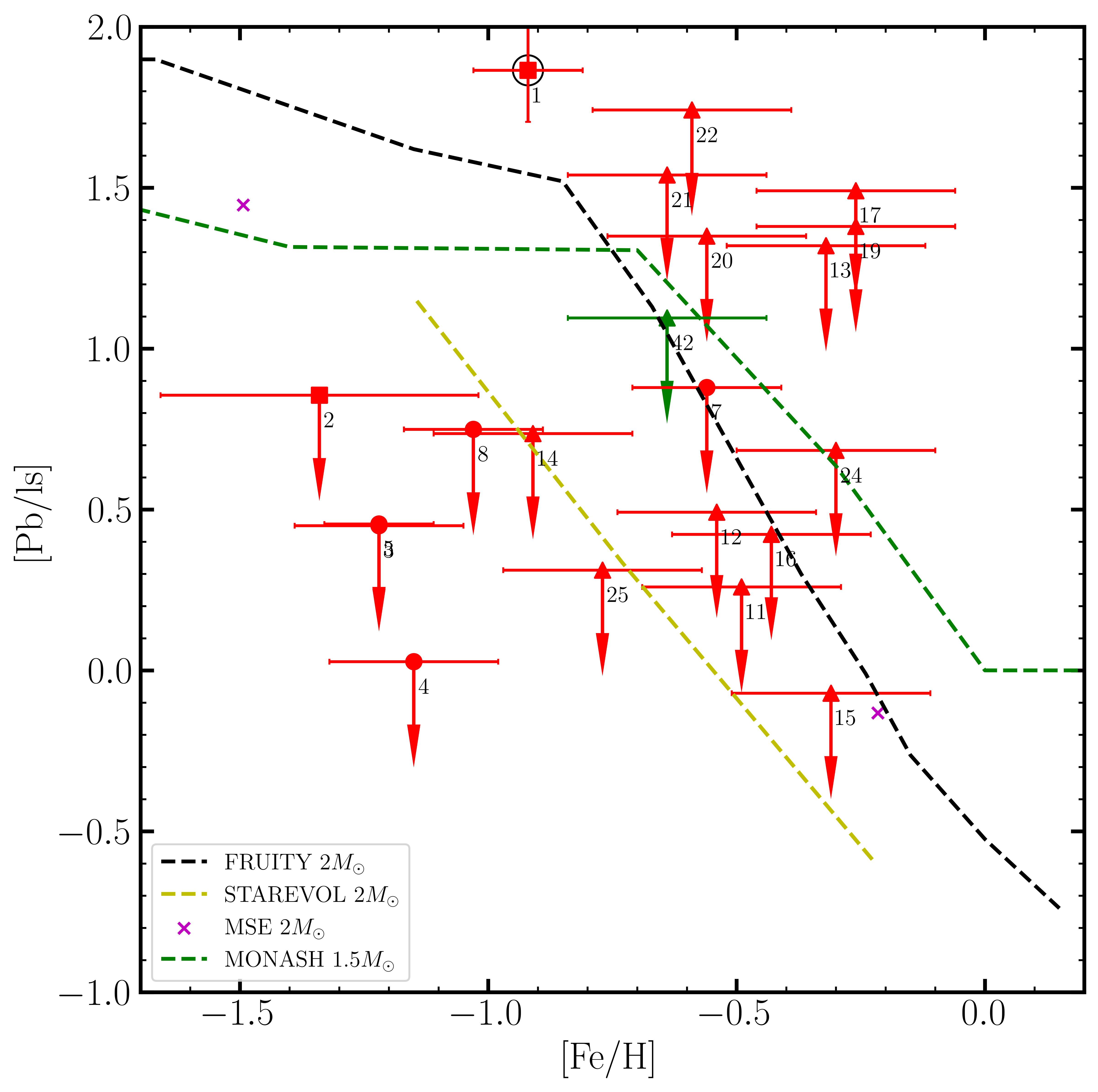}
    \end{subfigure}
    \caption{Same as Figure~\ref{fig:hsNLTE_lead}, but for \pbls\ versus \feh.}
    \label{fig:lsNLTE_lead}
\end{figure*}

To further investigate the impact of NLTE effects on Pb abundance determinations, we compare the derived \pbhs\ and \pbls\ upper limits by incorporating the derived NLTE corrections from the toy model for the Pb I lines. Figure~\ref{fig:hsNLTE_lead} and Figure~\ref{fig:lsNLTE_lead} present the \pbhs\ and \pbls\ upper limits (except for J003643) as a function of metallicity (\feh), respectively, adapted from \citet{desmedt2016}, for the comparative sample. These figures also include predictions from various stellar evolutionary models: $2~\Msun$ FRUITY model \citep{Cristallo2015}, $2~\Msun$ STAREVOL code \citep[][and references therein]{Goriely2004, Siess2007}, $2~\Msun$ Mount-Stromlo Evolutionary (MSE) predictions \citep[][and references therein]{fishlock14, karakas10}, and 1.5~~\Msun\ MONASH model \citep[see]{Karakas2014, karakas16, Karakas2018}. In both cases, the left figure shows LTE-based upper limits of Pb abundances, while the right figure includes NLTE-corrected upper limits of Pb abundances derived in this study. We note that NLTE corrections have been derived and applied exclusively to the upper limits of Pb abundances derived using Pb I lines, with no corrections applied to Pb II lines. For all stars except J003643, the \pbhs\ and \pbls\ ratios represent upper limits. The error bars along the x-axis indicate uncertainties in the metallicity (\feh) of the stars.

The comparison reveals that even after applying NLTE corrections, the majority of post-AGB stars at low metallicities (\feh\,$<-0.5$ dex) exhibit significantly lower upper limits of \pbhs\ and \pbls\ values than theoretical predictions. Figure~\ref{fig:hsNLTE_lead} shows that roughly half of the existing sample with \feh\,$<-0.5$ dex are compatible with the theoretical predictions within the error bars. This, however, may reflect observational bias because the current sample is far from complete. Although NLTE corrections tighten the upper limits of the Pb abundances, they do not fully resolve the disagreement at low metallicities. Additionally, at higher metallicities, the NLTE corrections lead to an overall increase in the derived upper limits of \pbhs\ and \pbls\ compared to their LTE-based values. This result suggests that NLTE effects must be accounted for in Pb abundance determinations, but they are insufficient to explain the observed discrepancies at low metallicities. Furthermore, J003643 is the first post-AGB star with a clear detection and precise derivation of Pb abundance, rather than an upper limit, which is significantly higher than that of the comparative sample (see Table~\ref{tab:lead}). Yet, even this robust detection remains markedly underpredicted by current nucleosynthesis models. This underscores persistent shortcomings in our theoretical understanding of \sprocess nucleosynthesis, particularly in the context of explaining Pb production at low metallicities.

\section{Summary and Conclusions}
\label{sec:conclusion}

In this paper, we expand the known chemical diversity among \sprocess enriched post-AGB stars, presenting a detailed abundance analysis of J003643, a single post-AGB star in the SMC. Using high-resolution VLT/UVES spectra and detailed analysis with E-iSpec, we derived the atmospheric parameters and elemental abundances of J003643. J003643 exhibits a C/O ratio of 16.21$\pm$0.22. With a \sfe\,= 2.09 dex, J003643 also provides the first direct Pb abundance measurement in a post-AGB star (\pbfe\,= 3.16$\pm$0.18 dex) using the Pb II 5608~\AA\ line, consistent across both EW and SSF methods.

To investigate the origin of J003643's relatively high C/O ratio, we compared its key abundance ratios with those of $\sim$28 comparative sample of post-AGB stars from the Galaxy, LMC, and SMC. While J003643 follows the expected \cfe--\sfe\ trend, consistent with a typical TDU history, it stands out as a clear outlier when compared in both the \ofe--\cfe\ and \ofe--\feh\ planes. J003643 exhibits a relatively low \ofe\,= –0.08$\pm$0.20 dex and [$\alpha$/Fe]\,$\approx$ 0 dex compared to its Galactic and LMC counterparts with similar \cfe\ and \feh, which typically show \ofe\,$\approx$ 0.5–1.0 dex. This relatively low \ofe and $\alpha$-abundance of J003643 is consistent with the chemical evolution of SMC at $\feh\,\approx$ –1 dex, and contrasts with the $\alpha$-enhanced trend observed in the Galaxy and LMC at the same metallicity. These comparisons suggest that the high C/O ratio in J003643 arises primarily from an oxygen-poor natal environment, rather than from an unusually high carbon enrichment.

Additionally, to investigate the nucleosynthetic origin of J003643’s heavy-element abundances, we first considered whether they could reflect its birth environment in the SMC. However, its abundance pattern differs markedly from that of typical SMC field giant stars, which are commonly used to trace the chemical evolution of the SMC. We then compared its abundances with predictions from various stellar evolutionary models: ATON, MONASH, and FRUITY, the latter two incorporating post-processing nucleosynthesis. While the models reproduce the observed abundance of CNO, $\alpha$, Fe-peak and \sprocess elements, they systematically underpredicts Pb abundance, despite the varying nucleosynthetic assumptions. This highlights persistent gaps in our understanding of lead production in AGB stars. 

To explore this, we analysed Pb abundances across the comparative sample of \sprocess enriched post-AGB stars using NLTE-corrected values from a simplified toy model. While NLTE corrections bring the upper limits a bit nearer to the model predictions, improving consistency in some cases, major discrepancies remain for stars with \feh$\,\lesssim$ -0.5 dex, further reinforcing our limited understanding of lead production.  

In summary, J003643 with its high C/O ratio and the first precise measurement of Pb abundance presents a valuable benchmark for advancing our understanding of CNO and \sprocess nucleosynthesis, providing critical constraints on AGB nucleosynthesis, dredge-up efficiency, and the development of improved model prescriptions for key heavy elements.

\begin{acknowledgement}
This study is based on observations collected with the Very Large Telescope at the ESO Paranal Observatory (Chile) of program number 092.D-0485. MM1 acknowledges the financial support provided by the International Macquarie Research Excellence Scholarship (iMQRES) program for the duration of this research. MM1, DK, and MM2 also acknowledge the ARC Centre of Excellence for All Sky Astrophysics in 3 Dimensions (ASTRO 3D). DK and HVW acknowledge the support of the Australian Research Council Discovery Project DP240101150. AMA acknowledges the support from the Swedish Research Council (VR 2020–03940) and from the Crafoord Foundation via the Royal Swedish Academy of Sciences (CR 2024–0015). SC acknowledges the funding by the European Union – NextGenerationEU RFF M4C2 1.1 PRIN 2022 project ‘2022RJLWHN URKA’ and the INAF Theory Grant ‘Understanding Rprocess \& Kilonovae Aspects (URKA)’. DV acknowledges the financial support from the German-Israeli Foundation (GIF No. I-1500-303.7/2019). AK acknowledges the support of the Australian Research Council Centre of Excellence for All Sky Astrophysics in 3 Dimensions (ASTRO 3D), through project number CE170100013. P.V. acknowledges the support received from the PRIN INAF 2019 grant ObFu 1.05.01.85.14 (“Building up the halo: chemodynamical tagging in the age of large surveys”, PI. S. Lucatello). HVW acknowledges support from the Research Council, KU Leuven, under grant numbers IRI-I000225N and I000325N. 
\end{acknowledgement}



\paragraph{Data Availability Statement}
The data underlying this article is made available online.

\printendnotes
\bibliography{mnemonic,meghna}

\appendix
\section{Linelist of J003643}
\label{sec:appendix:linelist}
\input{Linelist}

\section{Comparative Sample of Post-AGB Stars}
\label{sec:appendix:sample}
In this appendix, we present the comparative sample of post-AGB stars discussed in this study. Table~\ref{tab:full sample} provides a comprehensive list of single \sprocess enriched post-AGB stars, single non-\sprocess enriched post-AGB stars, and binary \sprocess enriched post-AGB stars, spanning different galactic environments, including the LMC, SMC, and the Galaxy. This dataset serves as the basis for our comparative analysis of heavy-element enrichment, neutron irradiation efficiency, and the dependence of chemical signatures on stellar parameters, as discussed in Section~\ref{sec:full_sample}.
\input{Table_Appendix}

\end{document}

%% file: Table1.tex
\begin{table}
    \caption{Photometric data used for the SED of J003643. See text for full details.}
    \begin{center}
            
        \begin{tabular}{ c c c}
            \hline
            \addlinespace
        Photometry band  & Survey & Magnitude \\
        \addlinespace
            \hline
            \addlinespace
            U & JOHNSON & 16.308 \\
            B & JOHNSON & 15.908 \\
            V & JOHNSON & 15.257 \\
            I & DENIS & 14.395 \\
            J & 2MASS & 13.942 \\
            H & 2MASS & 13.707 \\
            KS & 2MASS & 13.612 \\
            W1 & WISE & 13.465 \\
            W2 & WISE & 13.317 \\
            W3 & WISE & 8.331 \\
            W4 & WISE & 6.115 \\
            $[3.6]$ & IRAC & 13.373 \\
            $[4.5]$ & IRAC & 13.285 \\
            $[5.8]$ & IRAC & 11.464 \\
            $[8.0]$ & IRAC & 9.515 \\
            $[24]$ & MIPS & 6.139 \\
            \hline
        \end{tabular}
    \end{center}
    \begin{tablenotes}
     \small
    \item \textbf{Notes:} A standard error of 0.05 mag is adopted for the SED fitting. IRAC and MIPS are instruments aboard the Spitzer satellite.
    \end{tablenotes}
    \label{tab:photometry}    
\end{table}

%% file: Table2.tex
\begin{table*}
    \caption{Observational logs of the J003643.}
    \begin{center}
            
        \begin{tabular}{ c c c c c }
            \hline
            \addlinespace
        \begin{tabular}[c]{@{}c@{}}
        Date \\ {(}YYYY-MM-DD{)}\end{tabular} & \begin{tabular}[c]{@{}c@{}}
        UTC Start \\ {(}hh:mm:ss{)}\end{tabular} &  \begin{tabular}[c]{@{}c@{}}Exp Time\\ {(}s{)}\end{tabular} & \begin{tabular}[c]{@{}c@{}}
        Wavelength Coverage \\  {(}{\AA}{)}
        \end{tabular} & S/N \\
        \addlinespace
            \hline
            \addlinespace
             2013-10-19 & 02:03:19 & 2730 & 3282-4563 & 15\\  
            2013-10-19 & 02:53:07 & 2700 & 3282-4563 & 15\\ 
            2013-10-19 & 03:46:18 & 3000 & 3282-4563 & 18\\  
            2013-10-19 & 01:09:29 & 2900 & 3282-4563 & 14\\  
            2013-10-19 & 02:03:11 & 2730 & 4726-6835 & 52\\  
            2013-10-19 & 02:53:03 & 2700 & 4726-6835 & 52\\  
            2013-10-19 & 03:46:15 & 3000 & 4726-6835 & 60\\  
            2013-10-19 & 01:09:25 & 2900 & 4726-6835 & 52\\
            \hline
        \end{tabular}
    \end{center}
    \begin{tablenotes}
     \small
    \item \textbf{Notes:} The final S/N after the weighted mean merging of similar observations is approximately 30 in the blue region and 55 in the red region of the spectrum.
    \end{tablenotes}
    \label{tab:log}

\end{table*}

%% file: Table3.tex
\begin{table}
	\centering
	\caption{Spectroscopically derived atmospheric parameters of J003643.}
	\label{tab:atmosParam_J003643}
	\begin{tabular}{c c c} 
		\hline
            \addlinespace
             Stellar Parameters & This study & \citet{Kamath2014}\\
             \addlinespace
            \hline
            \addlinespace
		\Teff\ (K) & $7552\pm91$ & $7458\pm250$\\
		\logg\ (dex)  & $1.04\pm0.16$ & $0.50\pm0.50$\\
		\feh\ (dex) & $-0.92\pm0.11$ & $-0.63\pm0.50$\\
		\mv\ (km/s) & $4.02\pm0.13$ & $3.00\pm0.50$\\
            \addlinespace
            \hline
	\end{tabular}
    \begin{center}
     \begin{tablenotes}
     \small
    \item \textbf{Notes:} Atmospheric parameters from \citet{Kamath2014} were derived using low-resolution AAOmega spectra.
    \end{tablenotes}
    \end{center}
\end{table}

%% file: Table4.tex
\begin{table*}
    \centering
    \caption{Spectroscopically determined abundance results for J003643.}
        \label{tab:abund_J003643}
        \begin{tabular}{ c c c c c c c c c c }
            \hline
            \addlinespace
             Ion & Z & N & \logepsilonsun & \logepsilon & \xfelte & \sigmatotal & \xfenlte & eNLTE & \sigmalinetoline\\
             \hline
            \addlinespace
            C 1 & 6 & 14 & 8.43 & 8.90 & 1.39 & 0.14 & 1.33 & 0.14 & 0.12\\
            O 1 & 8 & 3 & 8.69 & 7.80 & 0.03 & 0.20 & -0.08 & 0.20 & 0.02\\
            Mg 1 & 12 & 2 & 7.60 & 6.66 & -0.01 & 0.13 & -0.09 & 0.13 & 0.03\\
            Ca 2 & 20 & 2 & 6.34 & 5.25 & -0.16 & 0.11 & -0.13 & 0.11 & 0.01\\
            Sc 2 & 21 & 2 & 3.15 & 2.07 & -0.16 & 0.10 &  &  & 0.13\\
            Ti 2 & 22 & 14 & 4.95 & 3.84 & -0.19 & 0.16 &  &  & 0.10\\
            Cr 1 & 24 & 2 & 5.64 & 4.85 & 0.13 & 0.21 &  &  & 0.13\\
            Cr 2 & 24 & 9 & 5.64 & 4.66 & -0.06 & 0.17 &  &  & 0.14\\
            Fe 1 & 26 & 46 & 7.50 & 6.56 & -0.02 & 0.17 &  &  & 0.11\\
            Fe 2 & 26 & 5 & 7.50 & 6.53 & -0.04 & 0.16 &  &  & 0.15\\
            Ni 1 & 28 & 1 & 6.22 & 5.28 & -0.02 & 0.22 &  &  & 0.20\\
            Y 2 & 39 & 14 & 2.21 & 2.89 & 1.60 & 0.19 &  &  & 0.17\\
            Zr 2 & 40 & 3 & 2.58 & 3.45 & 1.79 & 0.17 &  &  & 0.15\\
            La 2 & 57 & 18 & 1.10 & 2.62 & 2.44 & 0.17 &  &  & 0.15\\
            Ce 2 & 58 & 27 & 1.58 & 2.88 & 2.22 & 0.17 &  &  & 0.10\\
            Pr 2 & 59 & 6 & 0.72 & 2.18 & 2.38 & 0.15 &  &  & 0.12\\
            Nd 2 & 60 & 40 & 1.42 & 2.88 & 2.39 & 0.16 &  &  & 0.13\\
            Sm 2 & 62 & 5 & 0.96 & 2.16 & 2.12 & 0.12 &  &  & 0.04\\
            Eu 2 & 63 & 2 & 0.52 & 1.41 & 1.81 & 0.11 &  &  & 0.09\\
            Gd 2 & 64 & 2 & 1.07 & 2.05 & 1.90 & 0.18 &  &  & 0.07\\
            Er 2 & 68 & 2 & 0.92 & 2.33 & 2.33 & 0.21 &  &  & 0.11\\
            Yb 2 & 70 & 1 & 0.84 & 2.12 & 2.20 & 0.22 &  &  & 0.20\\
            Lu 2 & 71 & 1 & 0.10 & 1.25 & 2.07 & 0.21 &  &  & 0.20\\
            Hf 2 & 72 & 2 & 0.85 & 2.61 & 2.68 & 0.13 &  &  & 0.06\\
            Pb 2 & 82 & 1 & 1.75 & 3.98 & 3.16 & 0.21 &  &  & 0.20\\
            \addlinespace
            \hline
        \end{tabular}
    \begin{center}
    \begin{tablenotes}
     \small
    \item \textbf{Notes:} The ions that were detected and their corresponding atomic number (Z) are listed in columns 1 and 2, respectively. The solar abundances (\logepsilonsun) in column 3 are retrieved from \citet{Asplund2009}. N represents the number of lines used for each ion, \xfelte\ is the element-over-iron ratio derived considering LTE, \sigmatotal\ is the total uncertainty on \xfe, \logepsilon\ is the determined abundance, and \sigmalinetoline\ is the line-to-line scatter. We impose a \sigmalinetoline\ of 0.20 dex for all ions for which only one line is available for the abundance determination. \xfenlte\ is the element-over-iron ratio derived considering NLTE along with uncertainty eNLTE (see Section~\ref{sec:nlte} for details).
    \end{tablenotes}
    \end{center}  
\end{table*}

%% file: Table6.tex
\begin{table}
	\centering
	\caption{Overview of the effective temperature, metallicity, C/O ratio, \sprocess indices, derived luminosity and initial mass estimate of J003643.}
	\label{tab:summary}
	\begin{tabular}{c | c}
		\hline
            Parameter & Value\\
            \hline
            \Teff (K) & $7552 \pm 91$ \\
            \feh (dex) & $-0.97 \pm 0.11$ \\ 
            $\mathrm{C/O_{LTE}}$ &  $12.58 \pm 0.25$\\
            $\mathrm{C/O_{NLTE}}$ & $16.21 \pm 0.22$ \\
            \sfe & $2.09 \pm 0.20$ \\    
            \lsfe & 1.72 $\pm$ 0.17 \\ 
            \hsfe & 2.19 $\pm$ 0.11 \\
            \hsls & 0.48 $\pm$ 0.12\\
            \pbfe & 3.16 $\pm$ 0.18\\
            \pbls & 1.39 $\pm$ 0.16\\
            \pbhs & 1.86 $\pm$ 0.15\\
            \Lsed (\Lsun) & $7623^{8606}_{5761}$\\
		$E(B-V)$ (mag)  & $0.46^{0.50}_{0.36}$\\
            \hline
	\end{tabular}
    \begin{center}
     \begin{tablenotes}
     \small
    \item \textbf{Notes:} See Section~\ref{sec:abund_analysis} for details on chemical abundances presented here. See Section~\ref{sec:lum_sed} for details on luminosity derivation.
    \end{tablenotes}
    \end{center}
\end{table}

%% file: Table7.tex
\begin{sidewaystable}
\scriptsize
    \captionsetup{justification=raggedright,singlelinecheck=false}
    \caption{LTE and NLTE abundance ratios of lead (\pbls\ and \pbhs) for \sprocess enriched single and binary post-AGB stars in the Galaxy and MCs, where lead abundance measurements are available. The table also includes their effective temperature and metallicity. See text for details.}
    \resizebox{\linewidth}{!}{%
    \centering
        \begin{tabular}{ c c c c c c c c c c c c c}
            \hline
            \addlinespace
             Index & Object name & LMC/SMC/Galaxy & 
             \Teff & \feh & \logepsilon Pb I & \pbIfe & \logepsilon Pb II & \pbIIfe & \pbhs & \pbls & NLTE \pbhs & NLTE \pbls \\
             \hline
            \addlinespace
            1 &  J003643  &  SMC  &  7552 $\pm$ 91  &  -0.92 $\pm$ 0.11  &  ...  &  ...  & 4.29 & 3.58 & 1.39$\pm$0.15 & 1.865$\pm$0.16 &  ...  &  ... \\
            2 &  J004441.04-732136.4  &  SMC  &  6250 $\pm$ 125  &  -1.34 $\pm$ 0.32  &  <3.00  &  <2.58  &  ...  &  ...  &  <0.00  &  <0.52  &  <0.34  &  <0.86 \\
            3 &  J050632.10-714229.8  &  LMC  &  6750 $\pm$ 125  &  -1.22 $\pm$ 0.17  &  <2.05  &  <1.52  &  ...  &  ...  &  <0.45  &  <0.10  &  <0.80  &  <0.45 \\
            4 &  J052043.86-692341.0  &  LMC  &  5750 $\pm$ 125  &  -1.15 $\pm$ 0.17  &  <2.00  &  <1.40  &  ...  &  ...  &  <-0.45  &  <-0.27  &  <-0.15  &  <0.03 \\
            5 &  J053250.69-713925.8  &  LMC  &  5500 $\pm$ 125  &  -1.22 $\pm$ 0.11  &  <2.23  &  <1.70  &  ...  &  ...  &  <-0.24  &  <0.19  &  <0.03  &  <0.46 \\
            7 &  J051213.81-693537.1  &  LMC  &  5875 $\pm$ 125  &  -0.56 $\pm$ 0.15  &  <3.30  &  <2.11  &  ...  &  ...  &  <0.37  &  <0.78  &  <0.47  &  <0.88 \\
            8 &  J051848.86-700246.9  &  LMC  &  6000 $\pm$ 125  &  -1.03 $\pm$ 0.14  &  <2.62  &  <1.93  &  ...  &  ...  &  <-0.19  &  <0.47  &  <0.09  &  <0.75 \\
            11 &  IRAS 05113+1347  &  Galaxy  &  5500 $\pm$ 125  &  -0.49 $\pm$ 0.20  &  <2.65  &  <1.44  &  ...  &  ...  &  <-0.21  &  <0.11  &  <-0.06  &  <0.26 \\
            12 &  IRAS 05341+0852  &  Galaxy  &  6750 $\pm$ 125  &  -0.54 $\pm$ 0.20  &  <3.15  &  <2.10  &  ...  &  ...  &  <-0.14  &  <0.23  &  <0.12  &  <0.49 \\
            13 &  IRAS 06530-0213  &  Galaxy  &  7375 $\pm$ 125  &  -0.32 $\pm$ 0.20  &  <4.10  &  <2.73  &  <4.19  &  <2.76  &  <0.69  &  <0.98  &  <1.03  &  <1.32\\
            14 &  IRAS 07134+1005  &  Galaxy  &  7250 $\pm$ 125  &  -0.91 $\pm$ 0.20  &  <2.79  &  <2.01  &  ...  &  ...  &  <0.38  &  <0.37  &  <0.75  &  <0.74 \\
            15 &  IRAS 07430+1115  &  Galaxy  &  6000 $\pm$ 125  &  -0.31 $\pm$ 0.20  &  <2.50  &  <1.07  &  ...  &  ...  &  <-0.48  &  <-0.23  &  <-0.32  &  <-0.07 \\
            16 &  IRAS 08143-4406  &  Galaxy  &  7000 $\pm$ 125  &  -0.43 $\pm$ 0.20  &  <3.20  &  <1.90  &  ...  &  ...  &  <0.32  &  <0.13  &  <0.61  &  <0.42  \\
            17 &  IRAS 08281-4850  &  Galaxy  &  7875 $\pm$ 125  &  -0.26 $\pm$ 0.20  &  <4.20  &  <2.74  &  <4.20  &  <2.71  &  <1.13  &  <1.14  &  <1.48  &  <1.49 \\
            19 &  IRAS 13245-5036  &  Galaxy  &  9500 $\pm$ 125  &  -0.26 $\pm$ 0.20  &  ...  &  ...  &  <4.40  &  <2.94  &  <0.91  &  <1.38  &  ...  &  ... \\
            20 &  IRAS 14325-6428  &  Galaxy  &  8000 $\pm$ 125  &  -0.56 $\pm$ 0.20  &  ...  &  ...  &  <3.79  &  <2.6  &  <1.27  &  <1.35  &  ...  &  ...\\
            21 &  IRAS 14429-4539  &  Galaxy  &  9375 $\pm$ 125  &  -0.64 $\pm$ 0.20  &  ...  &  ...  &  <4.40  &  <2.83  &  <1.36  &  <1.54  &  ...  &  ... \\
            22 &  IRAS 19500-1709  &  Galaxy  &  8000 $\pm$ 125  &  -0.59 $\pm$ 0.20  &  <3.89  &  <2.72  &  ...  &  ...  &  <1.38  &  <1.35  &  <1.77  &  <1.74 \\
            24 &  IRAS 22223+4327  &  Galaxy  &  6500 $\pm$ 125  &  -0.3 $\pm$ 0.20  &  <3.25  &  <1.82  &  ...  &  ...  &  <0.94  &  <0.48  &  <1.14  &  <0.68\\
            25 &  IRAS 22272+5435  &  Galaxy  &  5750 $\pm$ 125  &  -0.77 $\pm$ 0.20  &  <2.70  &  <1.72  &  ...  &  ...  &  <-0.18  &  <0.11  &  <0.02  &  <0.31 \\
            43 &  HD 168616  &  Galaxy  &  7250 $\pm$ 125  &  -0.64 $\pm$ 0.20  &  <2.75  &  <1.51  &  ...  &  ...  &  <0.59  &  <0.77  &  <0.92  &  <1.10 \\
            \addlinespace
            \hline
        \end{tabular}
}
    \begin{tablenotes}
        \small
        \item \textbf{Notes:} All lead abundances and ratios, except for J003643, are upper limits derived using the SSF technique and have been adopted from \citet{desmedt2016} and \citet{Kamathuniverse2021}. NLTE-corrected Pb abundance ratios are presented in the last two columns. See text for details
    \end{tablenotes}
    \label{tab:lead}
\end{sidewaystable}

%% file: Linelist.tex
\begin{table}
    \centering
    \caption{Linelist of J003643}
        \label{tab:linelist_J003643}
        \begin{tabular}{ c c c c c c c c }
            \hline
            \addlinespace
            Ion & wavelength (\AA) & log gf & EP(eV) & EW (m\AA) \\
            \addlinespace
            \hline
            \addlinespace
            C 1 & 4812.917 & -3.377 & 7.480 & 15.50 \\
            C 1 & 4932.049 & -1.658 & 7.685 & 144.80 \\
            C 1 & 5017.086 & -2.457 & 7.946 & 43.90 \\
            C 1 & 5024.919 & -2.728 & 7.946 & 22.40 \\
            C 1 & 5039.057 & -1.790 & 7.946 & 100.40 \\
            C 1 & 5057.681 & -2.223 & 8.647 & 22.80 \\
            C 1 & 5153.569 & -2.187 & 8.647 & 31.00 \\
            C 1 & 5380.325 & -1.616 & 7.685 & 143.40 \\
            C 1 & 5547.265 & -2.253 & 8.640 & 20.80 \\
            C 1 & 6002.980 & -2.169 & 8.647 & 17.00 \\
            C 1 & 6007.173 & -2.062 & 8.640 & 20.40 \\
            ... & ... & ... & ... & ... \\
            \addlinespace
            \hline
        \end{tabular}
    \begin{center}
    \begin{tablenotes}
     \small
    \item \textbf{Notes:} Linelist used for the abundance determination of J003643. The last column shows the calculated equivalent width (EW). The full line list used for the spectroscopic analysis of J003643 is provided as online supporting material.
    \end{tablenotes}
    \end{center}  
\end{table}

%% file: Table_Appendix.tex
\begin{sidewaystable*}
    \captionsetup{justification=raggedright,singlelinecheck=false}
    \caption{Fundamental properties, chemical abundances, and \sprocess indices of the entire sample of \sprocess-rich single post-AGB stars, \sprocess non-enriched single post-AGB stars, and \sprocess-rich binary post-AGB stars in the Galaxy and MCs.}
    \resizebox{\linewidth}{!}{%
    \centering
    \begin{tabular}{ c c c c c c c c c c c c c c c c c c c c }
        \hline
        \addlinespace
        Index & Object Name & SMC/LMC/Galaxy & \Teff\ (K) & \logg\ (dex) & \mv\ (km/s) & \feh\ & C/O & [O/Fe] & [C/Fe] & [N/Fe] & [s/Fe] & [ls/Fe] & [hs/Fe] & [hs/ls] & L/\Lsun\ & $L/\Lsun_{upper\ limit}$ &  $L/\Lsun_{lower\ limit}$ & Initial Mass & Reference \\
        \addlinespace
        \hline
        \addlinespace
        \multicolumn{20}{c}{\textbf{Post-AGB single stars with \sprocess enrichment}} \\
        \addlinespace
        \hline
        \addlinespace
        1 &  J003643  &  SMC  &  7552 $\pm$ 91  &  1.04 $\pm$ 0.16  &  4.02 $\pm$ 0.13  &  -0.97 $\pm$ 0.11  &  16.21 $\pm$ 0.22  &  0.03 $\pm$ 0.12  &  1.39 $\pm$ 0.12  &  ...  &  2.09 $\pm$ 0.20  &  1.72 $\pm$ 0.17  &  2.19 $\pm$ 0.11  &  0.48 $\pm$ 0.12  & 7623 & 8606 & 5761 &  ~2.0 &  \\
        2 &  J004441.04-732136.4  &  SMC  &  6250 $\pm$ 125  &  0.50 $\pm$ 0.50  &  3.50 $\pm$ 0.50  &  -1.34 $\pm$ 0.32  &  1.90 $\pm$ 0.70  &  1.14 $\pm$ 0.50  &  1.67 $\pm$ 0.36  &  ...  &  2.40 $\pm$ 0.10  &  2.06 $\pm$ 0.10  &  2.58 $\pm$ 0.10  &  0.52 $\pm$ 0.10  & 7600 & 7800 & 7400 &  ~1.3  & 1,2 \\
        3 &  J050632.10-714229.9  &  LMC  &  6750 $\pm$ 125  &  0.50 $\pm$ 0.25  &  3.00 $\pm$ 0.25  &  -1.22 $\pm$ 0.17  &  1.50 $\pm$ 0.30  &  0.91 $\pm$ 0.24  &  1.16 $\pm$ 0.11  &  ...  &  1.19 $\pm$ 0.09  &  1.42 $\pm$ 0.07  &  1.07 $\pm$ 0.14  &  -0.35 $\pm$ 0.15  & 5400 & 6100 & 4700 &  ~1.5  & 1,3 \\
        4 &  J052043.86-692341.0  &  LMC  &  5750 $\pm$ 125  &  0.50 $\pm$ 0.50  &  3.00 $\pm$ 0.25  &  -1.15 $\pm$ 0.17  &  1.60 $\pm$ 0.90  &  0.86 $\pm$ 0.35  &  1.53 $\pm$ 0.20  &  ...  &  1.79 $\pm$ 0.08  &  1.67 $\pm$ 0.12  &  1.85 $\pm$ 0.10  &  0.19 $\pm$ 0.16  & 8700 & 9700 & 7700 &  ~1.5  & 1,3 \\
        5 &  J053250.69-713925.8  &  LMC  &  5500 $\pm$ 125  &  0.00 $\pm$ 0.25  &  3.00 $\pm$ 0.25  &  -1.22 $\pm$ 0.11  &  2.50 $\pm$ 0.70  &  0.88 $\pm$ 0.38  &  1.53 $\pm$ 0.21  &  ...  &  1.80 $\pm$ 0.08  &  1.51 $\pm$ 0.16  &  1.94 $\pm$ 0.10  &  0.43 $\pm$ 0.19  & 6500 & 7500 & 5500 &  ~1.5  & 1,3 \\
        6 &  J053253.51-695915.1  &  LMC  &  4750 $\pm$ 125  &  2.50 $\pm$ 0.25  &  2.50 $\pm$ 0.25  &  -0.54 $\pm$ 0.12  &  ...  &  ...  &  ...  &  ...  &  0.85 $\pm$ 0.07  &  0.55 $\pm$ 0.10  &  0.95 $\pm$ 0.13  &  0.39 $\pm$ 0.14  & 1400 & 300 & 1100 &  ~1.5  & 1,3 \\
        7 &  J051213.81-693537.1  &  LMC  &  5875 $\pm$ 125  &  1.00 $\pm$ 0.25  &  3.00 $\pm$ 0.20  &  -0.56 $\pm$ 0.15  &  1.30 $\pm$ 0.40  &  0.52 $\pm$ 0.26  &  0.88 $\pm$ 0.26  &  ...  &  1.61 $\pm$ 0.06  &  1.33 $\pm$ 0.08  &  1.74 $\pm$ 0.08  &  0.41 $\pm$ 0.12  & 6700 & 6900 & 6500 &  1-1.5  & 1,4 \\
        8 &  J051848.86-700246.9  &  LMC  &  6000 $\pm$ 125  &  0.50 $\pm$ 0.25  &  2.80 $\pm$ 0.20  &  -1.03 $\pm$ 0.14  &  1.30 $\pm$ 0.30  &  0.84 $\pm$ 0.19  &  1.21 $\pm$ 0.16  &  ...  &  1.90 $\pm$ 0.07  &  1.46 $\pm$ 0.10  &  2.12 $\pm$ 0.08  &  0.66 $\pm$ 0.13  & 6250 & 6450 & 6050 &  1-1.5  & 1,4 \\
        9 &  IRAS Z02229+6208  &  Galaxy  &  5500 $\pm$ 250  &  0.50 $\pm$ 0.25  &  4.25 $\pm$ 0.25  &  -0.45 $\pm$ 0.14  &  ...  &  ...  &  0.78 $\pm$ 0.15  &  1.19 $\pm$ 0.30  &  1.40 $\pm$ 0.15  &  2.16 $\pm$ 0.12  &  1.29 $\pm$ 0.03  &  -0.87 $\pm$ 0.20  & 12959 & 16911 & 9973 &  3-3.5  & 1,5 \\
        10 &  IRAS 04296+3429  &  Galaxy  &  7000 $\pm$ 250  &  1.00 $\pm$ 0.25  &  4.00 $\pm$ 0.25  &  -0.62 $\pm$ 0.11  &  ...  &  ...  &  0.80 $\pm$ 0.20  &  0.39 $\pm$ 0.01  &  1.50 $\pm$ 0.23  &  1.70 $\pm$ 0.23  &  1.50 $\pm$ 0.17  &  -0.20 $\pm$ 0.23  & 10009 & 20082 & 5971 &  1-1.5  & 1,6 \\
        11 &  IRAS 05113+1347  &  Galaxy  &  5500 $\pm$ 125  &  0.50 $\pm$ 0.25  &  5.00 $\pm$ 0.25  &  -0.49 $\pm$ 0.17  &  2.42 $\pm$ 0.40  &  0.01 $\pm$ 0.27  &  0.65 $\pm$ 0.16  &  ...  &  1.54 $\pm$ 0.07  &  1.33 $\pm$ 0.13  &  1.65 $\pm$ 0.07  &  0.32 $\pm$ 0.15  & 2037 & 6731 & 1043 &  1-1.3  & 1,7 \\
        12 &  IRAS 05341+0852  &  Galaxy  &  6750 $\pm$ 125  &  1.25 $\pm$ 0.25  &  3.50 $\pm$ 0.25  &  -0.54 $\pm$ 0.11  &  1.06 $\pm$ 0.30  &  0.75 $\pm$ 0.11  &  1.03 $\pm$ 0.10  &  ...  &  2.12 $\pm$ 0.05  &  1.87 $\pm$ 0.08  &  2.24 $\pm$ 0.06  &  0.37 $\pm$ 0.10  & 324 & 592 & 197 &  0.5-0.6  & 1,7 \\
        13 &  IRAS 06530-0213  &  Galaxy  &  7375 $\pm$ 125  &  1.25 $\pm$ 0.25  &  4.00 $\pm$ 0.25  &  -0.32 $\pm$ 0.11  &  1.66 $\pm$ 0.39  &  0.35 $\pm$ 0.11  &  0.83 $\pm$ 0.13  &  ...  &  1.94 $\pm$ 0.06  &  1.75 $\pm$ 0.09  &  2.04 $\pm$ 0.08  &  0.29 $\pm$ 0.13  & 4687 & 8178 & 2736 &  1.5-2  & 1,7 \\
        14 &  IRAS 07134+1005  &  Galaxy  &  7250 $\pm$ 125  &  0.50 $\pm$ 0.25  &  3.25 $\pm$ 0.25  &  -0.91 $\pm$ 0.20  &  1.24 $\pm$ 0.29  &  0.81 $\pm$ 0.19  &  1.16 $\pm$ 0.22  &  0.57 $\pm$ 0.19  &  1.63 $\pm$ 0.14  &  1.64 $\pm$ 0.13  &  1.63 $\pm$ 0.20  &  -0.01 $\pm$ 0.24  & 5505 & 6098 & 4955 &  0.9-1.2  & 1,7 \\
        15 &  IRAS 07430+1115  &  Galaxy  &  6000 $\pm$ 125  &  1.00 $\pm$ 0.25  &  3.25 $\pm$ 0.25  &  -0.35 $\pm$ 0.15  &  1.71 $\pm$ 0.30  &  0.30 $\pm$ 0.22  &  0.79 $\pm$ 0.13  &  ...  &  1.47 $\pm$ 0.06  &  1.30 $\pm$ 0.14  &  1.55 $\pm$ 0.06  &  0.25 $\pm$ 0.15  & 20 & 30 & 14 &  0.5-0.6  & 1,7 \\
        16 &  IRAS 08143-4406  &  Galaxy  &  7000 $\pm$ 125  &  1.50 $\pm$ 0.25  &  5.50 $\pm$ 0.25  &  -0.43 $\pm$ 0.11  &  1.66 $\pm$ 0.39  &  0.19 $\pm$ 0.13  &  0.67 $\pm$ 0.12  &  0.01 $\pm$ 0.22  &  1.65 $\pm$ 0.05  &  1.77 $\pm$ 0.08  &  1.58 $\pm$ 0.06  &  -0.19 $\pm$ 0.11  & 4509 & 5452 & 3927 &  1-1.5  & 1,7 \\
        17 &  IRAS 08281-4850  &  Galaxy  &  7875 $\pm$ 125  &  1.25 $\pm$ 0.25  &  5.50 $\pm$ 0.25  &  -0.26 $\pm$ 0.11  &  2.34 $\pm$ 0.42  &  0.12 $\pm$ 0.11  &  0.75 $\pm$ 0.21  &  ...  &  1.58 $\pm$ 0.09  &  1.57 $\pm$ 0.11  &  1.58 $\pm$ 0.12  &  0.01 $\pm$ 0.17  & 9584 & 16692 & 5567 &  1.5-2  & 1,7 \\
        18 &  IRAS 12360–5740  &  Galaxy  &  7273 $\pm$ 250  &  1.59 $\pm$ 0.25  &  3.00 $\pm$ 0.50  &  -0.40 $\pm$ 0.15  &  0.45 $\pm$ 0.20  &  0.31 $\pm$ 0.05  &  0.27 $\pm$ 0.18  &  0.22 $\pm$ 0.32  &  1.88 $\pm$ 0.20  &  1.73 $\pm$ 0.20  &  2.02 $\pm$ 0.20  &  0.29 $\pm$ 0.20  & 6265 & 7940 & 5178 &  1-1.5  & 1,8 \\
        19 &  IRAS 13245-5036  &  Galaxy  &  9500 $\pm$ 125  &  2.75 $\pm$ 0.25  &  4.50 $\pm$ 0.25  &  -0.30 $\pm$ 0.10  &  1.11 $\pm$ 0.30  &  0.26 $\pm$ 0.13  &  0.57 $\pm$ 0.21  &  ...  &  1.88 $\pm$ 0.09  &  1.56 $\pm$ 0.14  &  2.03 $\pm$ 0.11  &  0.47 $\pm$ 0.18  & 11221 & 16800 & 7106 &  1.5-2  & 1,7 \\
        20 &  IRAS 14325-6428  &  Galaxy  &  8000 $\pm$ 125  &  1.00 $\pm$ 0.25  &  5.75 $\pm$ 0.25  &  -0.56 $\pm$ 0.10  &  2.27 $\pm$ 0.40  &  0.57 $\pm$ 0.09  &  1.18 $\pm$ 0.23  &  0.18 $\pm$ 0.20  &  1.30 $\pm$ 0.14  &  1.25 $\pm$ 0.15  &  1.33 $\pm$ 0.19  &  0.08 $\pm$ 0.24  & 4935 & 6988 & 3758 &  1.5-2  & 1,7 \\
        21 &  IRAS 14429-4539  &  Galaxy  &  9375 $\pm$ 125  &  2.50 $\pm$ 0.25  &  4.75 $\pm$ 0.25  &  -0.18 $\pm$ 0.11  &  1.29 $\pm$ 0.26  &  0.31 $\pm$ 0.12  &  0.68 $\pm$ 0.23  &  ...  &  1.41 $\pm$ 0.08  &  1.29 $\pm$ 0.15  &  1.47 $\pm$ 0.10  &  0.18 $\pm$ 0.08  & 5049 & 14624 & 1591 &  1.5-2  & 1,7 \\
        22 &  IRAS 19500-1709  &  Galaxy  &  8000 $\pm$ 125  &  1.00 $\pm$ 0.25  &  6.00 $\pm$ 0.25  &  -0.59 $\pm$ 0.10  &  1.02 $\pm$ 0.17  &  0.72 $\pm$ 0.09  &  0.99 $\pm$ 0.33  &  0.41 $\pm$ 0.17  &  1.35 $\pm$ 0.21  &  1.37 $\pm$ 0.29  &  1.34 $\pm$ 0.30  &  -0.03 $\pm$ 0.41  & 7053 & 8183 & 6194 &  1.5-2  & 1,7 \\
        23 &  IRAS 20000+3239  &  Galaxy  &  5000 $\pm$ 250  &  ...  &  9.00 $\pm$ 0.25  &  -1.40 $\pm$ 0.20  &  ...  &  ...  &  1.70 $\pm$ 0.20  &  2.10 $\pm$ 0.20  &  1.40 $\pm$ 0.20  &  1.10 $\pm$ 0.20  &  1.47 $\pm$ 0.10  &  0.34 $\pm$ 0.20  & 14342 & 25218 & 9332 &  3-3.5  & 1,9 \\
        24 &  IRAS 22223+4327  &  Galaxy  &  6500 $\pm$ 125  &  1.00 $\pm$ 0.25  &  4.75 $\pm$ 0.25  &  -0.31 $\pm$ 0.20  &  1.04 $\pm$ 0.22  &  0.31 $\pm$ 0.13  &  0.59 $\pm$ 0.07  &  0.15 $\pm$ 0.21  &  1.03 $\pm$ 0.05  &  1.34 $\pm$ 0.07  &  0.88 $\pm$ 0.07  &  -0.46 $\pm$ 0.10  & 2163 & 2499 & 1956 &  0.5-0.6  & 1,7 \\
        25 &  IRAS 22272+5435  &  Galaxy  &  5750 $\pm$ 125  &  0.50 $\pm$ 0.25  &  4.25 $\pm$ 0.25  &  -0.77 $\pm$ 0.20  &  1.46 $\pm$ 0.26  &  0.63 $\pm$ 0.09  &  1.05 $\pm$ 0.12  &  ...  &  1.80 $\pm$ 0.05  &  1.61 $\pm$ 0.08  &  1.90 $\pm$ 0.07  &  0.28 $\pm$ 0.11  & 5659 & 6108 & 5234 &  1-1.3  & 1,7 \\
        26 &  IRAS 23304+6147  &  Galaxy  &  6750 $\pm$ 125  &  0.50 $\pm$ 0.25  &  3.00 $\pm$ 0.25  &  -1.00 $\pm$ 0.20  &  2.90 $\pm$ 0.30  &  0.20 $\pm$ 0.10  &  0.90 $\pm$ 0.12  &  0.50 $\pm$ 0.20  &  1.60 $\pm$ 0.07  &  1.55 $\pm$ 0.10  &  1.63 $\pm$ 0.09  &  0.09 $\pm$ 0.24  & 7712 & 9386 & 6831 &  2-2.5  & 1,10 \\
        \addlinespace
        \hline
        \addlinespace
        \multicolumn{20}{c}{\textbf{Post-AGB single stars without \sprocess enrichment}} \\
        \addlinespace
        \hline
        \addlinespace
        27 &  IRAS 01259+6823  &  Galaxy  &  5000 $\pm$ 125  &  1.50 $\pm$ 0.25  &  3.30 $\pm$ 0.25  &  -0.60 $\pm$ 0.10  &  0.40 $\pm$ 0.30  &  0.31 $\pm$ 0.06  &  0.18 $\pm$ 0.30  &  ...  &  0.30 $\pm$ 0.10  &  ...  &  ...  &  ...  & 340 & 646 & 220 &  0.5-0.6  & 1,11 \\
        28 &  IRAS 08187–1905  &  Galaxy  &  6250 $\pm$ 125  &  0.50 $\pm$ 0.25  &  3.80 $\pm$ 0.25  &  -0.60 $\pm$ 0.10  &  ...  &  0.26 $\pm$ 0.10  &  0.62 $\pm$ 0.30  &  0.49 $\pm$ 0.30  &  ...  &  ...  &  ...  &  ...  & 2619 & 3286 & 2099 &  0.5-0.6  & 1,11 \\
        29 &  SAO 239853  &  Galaxy  &  7452 $\pm$ 250  &  1.49 $\pm$ 0.25  &  5.00 $\pm$ 0.25  &  -0.81 $\pm$ 0.10  &  ...  &  0.80 $\pm$ 0.20  &  0.40 $\pm$ 0.20  &  0.60 $\pm$ 0.20  &  -0.40 $\pm$ 0.20  &  ...  &  ...  &  ...  & 23490 & 48520 & 13080 &  4-5  & 1,12 \\
        30 &  HD 107369  &  Galaxy  &  7533 $\pm$ 250  &  2.45 $\pm$ 0.25  &  2.50 $\pm$ 0.25  &  -1.10 $\pm$ 0.10  &  ...  &  0.00 $\pm$ 0.20  &  <-0.2  &  0.49 $\pm$ 0.30  &  -0.10 $\pm$ 0.20  &  ...  &  ...  &  ...  & 910 & 1010 & 814 &  0.5-0.6  & 1,12 \\
        31 &  HD 112374  &  Galaxy  &  6393 $\pm$ 250  &  0.80 $\pm$ 0.25  &  3.00 $\pm$ 0.50  &  -1.20 $\pm$ 0.10  &  ...  &  0.80 $\pm$ 0.20  &  0.10 $\pm$ 0.20  &  0.50 $\pm$ 0.20  &  -0.30 $\pm$ 0.20  &  ...  &  ...  &  ...  & 10777 & 11882 & 9961 &  2.5-3  & 1,12 \\
        32 &  HD 133656  &  Galaxy  &  8283 $\pm$ 250  &  1.38 $\pm$ 0.25  &  3.00 $\pm$ 0.50  &  -0.70 $\pm$ 0.10  &  ...  &  0.60 $\pm$ 0.20  &  0.30 $\pm$ 0.20  &  0.50 $\pm$ 0.20  &  -0.40 $\pm$ 0.20  &  ...  &  ...  &  ...  & 5227 & 5690 & 4861 &  0.8-1  & 1,12 \\
        33 &  HR 6144  &  Galaxy  &  6728 $\pm$ 250  &  0.93 $\pm$ 0.25  &  3.00 $\pm$ 0.50  &  -0.40 $\pm$ 0.10  &  ...  &  0.30 $\pm$ 0.20  &  0.30 $\pm$ 0.20  &  0.90 $\pm$ 0.20  &  0.20 $\pm$ 0.20  &  ...  &  ...  &  ...  & 25491 & 30419 & 22212 &  4-5  & 1,12 \\
        34 &  HD 161796  &  Galaxy  &  6139 $\pm$ 250  &  0.99 $\pm$ 0.25  &  3.00 $\pm$ 0.50  &  -0.30 $\pm$ 0.10  &  ...  &  0.40 $\pm$ 0.20  &  0.30 $\pm$ 0.20  &  1.10 $\pm$ 0.20  &  0.00 $\pm$ 0.20  &  ...  &  ...  &  ...  & 5742 & 6322 & 5209 &  1-1.2  & 1,12 \\
        35 &  IRAS 18025-3906  &  Galaxy  &  6154 $\pm$ 250  &  1.18 $\pm$ 0.25  &  3.00 $\pm$ 0.50  &  -0.51 $\pm$ 0.15  & 0.43 &  0.56 $\pm$ 0.20  &  0.46 $\pm$ 0.20  &  0.74 $\pm$ 0.20  &  ...  &  ...  &  ...  &  ...  & 2324 & 12963 & 975 &  0.8-1  & 1,13 \\
        36 &  HD 335675  &  Galaxy  &  6082 $\pm$ 250  &  1.58 $\pm$ 0.25  &  3.00 $\pm$ 0.50  &  -0.90 $\pm$ 0.20  & 0.25 &  0.77 $\pm$0.19  & 0.40 $\pm$ 0.20 &  <0.27  &  ...  &  ...  &  ...  &  ...  & 15843 & 28359 & 7303 &  ...  & 1,14 \\
        37 &  IRAS 19386+0155  &  Galaxy  &  6303 $\pm$ 250  &  1.00 $\pm$ 0.25  &  3.00 $\pm$ 0.50  &  -1.10 $\pm$ 0.14  &  ...  &  ...  &  0.10 $\pm$ 0.20  &  ...  &  -0.30 $\pm$ 0.20  &  ...  &  ...  &  ...  & 9611 & 22765 & 4345 &  0.7-0.8  & 1,15 \\
        38 &  IRAS 19475+3119  &  Galaxy  &  8216 $\pm$ 250  &  1.01 $\pm$ 0.25  &  3.00 $\pm$ 0.50  &  -0.24 $\pm$ 0.15  & 0.19 &  0.30 $\pm$ 0.02  &  -0.09 $\pm$ 0.30  &  ...  &  -0.30 $\pm$ 0.10  &  ...  &  ...  &  ...  & 6775 & 7545 & 5955 &  0.8-1  & 1,16 \\
        39 &  HR 7671  &  Galaxy  &  6985 $\pm$ 250  &  0.83 $\pm$ 0.25  &  3.00 $\pm$ 0.50  &  -1.60 $\pm$ 0.10  & 0.05 &  0.46 $\pm$ 0.05  &  -0.57 $\pm$ 0.13  &  0.51 $\pm$ 0.16  &  ...  &  ...  &  ...  &  ...  & 3579 & 3734 & 3449 &  0.5-0.6  & 1,17 \\
        40 & J005252.87-722842.9 & SMC & 8500$\pm$125 & 1.50$\pm$0.25 & 2$\pm$0.25 & -1.20$\pm$0.15 & ... &  0.70$\pm$0.10 & ... & ... & ... & ... & ... & ... &  8200 & 8400 & 8000 & 1.5-2.0 & 18\\
         \addlinespace
        \hline
        \addlinespace
        \multicolumn{20}{c}{\textbf{Post-AGB binary stars with \sprocess enrichment}} \\
        \addlinespace
        \hline
        \addlinespace
        41 &  J005107.19-734133.3  &  SMC  &  5768 $\pm$ 85  &  0.21 $\pm$ 0.25  &  4.08 $\pm$ 0.05  &  -1.57 $\pm$ 0.10  &  0.58 $\pm$ 0.37  &  1.06 $\pm$ 0.21  &  1.09 $\pm$ 0.21  &  ...  &  1.52 $\pm$ 0.20  &  0.94 $\pm$ 0.12  &  1.59 $\pm$ 0.10  &  0.65 $\pm$ 0.13  & 2868 & 6469 & 1405 &  0.8-1  & 19 \\
        42 &  MACHO 47.2496.8  &  LMC  &  4900 $\pm$ 250  &  0.00 $\pm$ 0.50  &  3.00 $\pm$ 1.00  &  -1.50 $\pm$ 0.50  &  >2  &  ...  &  ...  &  <0.88 $\pm$ 0.20  &  1.96 $\pm$ 0.20  &  1.21 $\pm$ 0.17  &  2.06 $\pm$ 0.12  &  0.85 $\pm$ 0.12  & 2208 & 4087 & 1877 &  1-1.2  & 19 \\
        43 &  HD 158616  &  Galaxy  &  7250 $\pm$ 250  &  1.25 $\pm$ 0.25  &  3.00 $\pm$ 0.50  &  -0.64 $\pm$ 0.12  &  0.94 $\pm$ 0.22  &  0.24 $\pm$ 0.16  &  0.51 $\pm$ 0.20  &  -0.07 $\pm$ 0.35  &  0.96 $\pm$ 0.17  &  0.74 $\pm$ 0.12  &  0.92 $\pm$ 0.11  &  0.18 $\pm$ 0.10  & 8256 & 10301 & 6288 &  ~2.5  & 19 \\              
        \addlinespace
        \hline
    \end{tabular}
}
    \begin{tablenotes}
        \small
        \item \textbf{Notes:} References: 1 - \citet{Kamath2022}, 2 - \citet{desmedt12}, 3 - \citet{Vanaarle2013}, 4 - \cite{deSmedt2015}, 5 - \citet{reddy99}, 6 - \citet{Vanwinckel2000}, 7 - \citet{desmedt2016}, 8 - \citet{pereira11}, 9 - \citet{klochkova06}, 10 - \cite{ReyniersThesis}, 11 - \citet{rao12}, 12 - \citet{vanwinckel97a}, 13 - \citet{molina19}, 14 - \citet{sahin11}, 15 - \citet{pereira04}, 16 - \citet{arellanoferro01}, 17 - \cite{reyniers05}, 18 - \citet{kamath17}, 19 - \citet{Menon2024}.
    \end{tablenotes}
    \label{tab:full sample}
\end{sidewaystable*}